\documentclass[12pt, draftclsnofoot, onecolumn]{IEEEtran}
\usepackage[dvips]{graphicx}
\usepackage{subfigure}
\usepackage{textcomp,amssymb,amsmath,dsfont,amsthm}
\usepackage{color}
\usepackage{cite}
\usepackage{algorithm}

\newtheorem{theorem}{Theorem}
\newtheorem{lemma}[theorem]{Lemma}

\def\bk{{\mathbf{k}}}
\def\bm{{\mathbf{m}}}
\def\bo{{\mathbf{o}}}
\def\bq{{\mathbf{q}}}

\def\bu{{\mathbf{u}}}
\def\bv{{\mathbf{v}}}
\def\bx{{\mathbf{x}}}
\def\by{{\mathbf{y}}}
\def\bz{{\mathbf{z}}}

\def\bC{{\mathbf{C}}}
\def\bD{{\mathbf{D}}}
\def\bG{{\mathbf{G}}}
\def\bU{{\mathbf{U}}}
\def\bQ{{\mathbf{Q}}}

\def\calC{{\mathcal{C}}}
\def\calF{{\mathcal{F}}}
\def\calI{{\mathcal{I}}}
\def\calL{{\mathcal{L}}}
\def\calN{{\mathcal{N}}}
\def\calO{{\mathcal{O}}}
\def\calQ{{\mathcal{Q}}}
\def\calV{{\mathcal{V}}}

\def\bPhi{{\boldsymbol{\Phi}}}
\def\bTheta{{\boldsymbol{\Theta}}}
\def\balpha{{\boldsymbol{\alpha}}}
\def\bgamma{{\boldsymbol{\gamma}}}

\def\blambda{{\boldsymbol{\lambda}}}
\def\bLambda{{\boldsymbol{\Lambda}}}

\def\tr{{\rm tr}}
\def\diag{{\rm diag}}

\def\black{\color{black}}

\definecolor{dark green}{rgb}{0.0, 0.5, 0.0}

\begin{document}
\title{Optimized Random Deployment of Energy\\[-.4cm] Harvesting Sensors for Field Reconstruction in\\[-.4cm] Analog and Digital Forwarding Systems\\[-.2cm]}
\author{Teng-Cheng Hsu, Y.-W. Peter Hong, and Tsang-Yi Wang \thanks{Y.-W. Peter Hong is the corresponding author. T.-C. Hsu and Y.-W. P. Hong (emails: {\tt tchsu@erdos.ee.nthu.edu.tw} and {\tt ywhong@ee} {\tt .nthu.edu.tw}) are with the Institute of Communications Engineering, National Tsing Hua University, Hsinchu, Taiwan. T.-Y. Wang (email: {\tt tcwang@faculty.nsysu.edu.tw}) is with the Institute of Communications Engineering, National Sun Yat-sen University, Kaohsiung, Taiwan.}

\thanks{This work was supported in part by the Ministry of Science and Technology, Taiwan, under grant 102-2221-E-007-016-MY3.}}

% The paper headers
%\markboth{Journal of \LaTeX\ Class Files,~Vol.~11, No.~4, December~2012}%
%{Shell \MakeLowercase{\textit{et al.}}: Bare Demo of IEEEtran.cls for Journals}

% make the title area
\maketitle
\vspace{-1.4cm}

{\renewcommand{\baselinestretch}{1.35}
\begin{abstract}
This work examines the large-scale deployment of energy harvesting sensors for the purpose of sensing and reconstruction of a spatially correlated Gaussian random field. The sensors are powered solely by energy harvested from the environment and are deployed randomly according to a spatially non-homogeneous Poisson point process whose {\black density} depends on the energy arrival statistics at different locations. Random deployment is suitable for applications that require deployment over a wide and/or hostile area. During an observation period, each sensor takes a local sample of the random field and reports the data to the closest data-gathering node if sufficient energy is available for transmission. The realization of the random field is then reconstructed at the fusion center based on the reported sensor measurements. For the purpose of field reconstruction, the sensors should, on the one hand, be more spread out over the field to gather more informative samples, but should, on the other hand, be more concentrated at locations with high energy arrival rates or large channel gains toward the closest data-gathering node. This tradeoff is exploited in the optimization of the random sensor deployment in both analog and digital forwarding systems. More specifically, given the statistics of the energy arrival at different locations and a constraint on the average number of sensors, the spatially-dependent sensor {\black density} and the energy-aware transmission policy at the sensors are determined for both cases by minimizing an upper bound on the average mean-square reconstruction error. The efficacy of the proposed schemes are demonstrated through numerical simulations.
\end{abstract}
}

%%%%%%%%%%%%%%%%%%%
%% Introdunction  %
%%%%%%%%%%%%%%%%%%%
\section{Introduction}
Wireless sensor networks (WSNs) consist of spatially distributed sensors that have the ability to sense the physical environment, process the gathered information, and communicate through the wireless interface. In recent years, WSNs have been adopted in a wide range of applications, such as environmental monitoring, disaster recovery, and battlefield surveillance, etc \cite{Akyildiz2002}. In these applications, sensors are often deployed in large-scale and in hostile areas making human maintenance and battery-replacement impractical. Due to these reasons, energy harvesting techniques for sensor nodes \cite{Raghunathan2006, Seah2009, Sudevalayam2011}, which enable the conversion of ambient energy (such as solar \cite{Taneja2008}, vibrational \cite{Zhu2012}, or thermal energy \cite{Paradiso2005}) to electric energy, are being developed and used to prolong sensor lifetime.
By employing energy harvesting technology, the characteristics of the energy arrival and the efficiency of energy usage will have a significant impact on the sensing performance. It is therefore necessary to adapt the sensor deployment and sensor operations to spatial variations of the energy arrival process.

The main objective of this work {\black is to determine optimal sensor deployment strategies, namely, spatial densities of energy harvesting sensors}, for the purpose of sensing and reconstruction of a spatially correlated Gaussian random field. The sensors are assumed to be deployed randomly and in large scale according to a spatially non-homogeneous Poisson point process (NHPPP) \cite{Baccelli2009, Wang2011, Streit2010}. During an observation period, each sensor takes a local sample of the random field and reports the observation to the closest data-gathering node based on a threshold-based energy-aware transmission control policy. The policy allows the sensor to transmit only if it has accumulated enough energy for transmission. We consider both analog-forwarding (AF) systems, where sensors transmit a scaled version of their analog measurements to the data-gathering node, and digital-forwarding (DF) systems, where digital representations of their measurements are forwarded instead. The random field is then reconstructed at the fusion center based on the information gathered from the sensors.

In this work, we assume that the sensors' operations are supported solely by ambient energy (i.e., energy harvested from the environment) and, thus, their transmit powers and probabilities depend strongly on the characteristics of their energy arrival at their respective locations. To reduce the field reconstruction error, the sensors should, on the one hand, be more spread out over the field to gather more informative samples, but should, on the other hand, be concentrated more at locations with large energy arrivals or with large channel gains toward the closest data-gathering node. This tradeoff is exploited to determine the optimal random sensor deployment in both AF and DF systems. In particular, given {\black the locations of data-gathering nodes\footnote{\black Here, we assume that the locations of the data-gathering nodes are fixed and focus on the deployment of energy harvesting sensors. This is often the case in practice since data-gathering nodes are typically grid-connected and, thus, their placement can be more restrictive. However, in certain cases, further flexibility may also be given to the deployment of data-gathering nodes. Readers are referred to \cite{Kwon2013} for further discussions on this topic.},} the statistics of the energy arrival at different locations and a constraint on the average number of sensors, we determine the spatially-dependent sensor {\black densities} and the energy-aware transmission thresholds at the sensors by minimizing an upper bound on the average mean-square reconstruction error. The energy-aware transmission policy allows a sensor to transmit only when its accumulated energy (i.e., its battery level) is beyond a certain threshold and remains silent, otherwise, so that energy can be preserved for use in later time slots.
%is implemented in accordance with each sensor's accumulated energy level, based on which each sensor can determine optimally whether to
%transmit in a given observation period or remain silent so that energy is preserved for later time periods.
Notice that, {\black in this work, we consider only a simple threshold-based transmission policy and focus on the global effect of the sensor deployment problem.
%do not focus on more complicated designs of the transmission control policies, as done in \cite{Yang2010,Ozel2011,Sharma2010}, but focus on the sensor deployment problem using a simple threshold-based policy.
In general, the optimal energy-aware transmission policy may involve continuous power control in accordance with the data traffic, the transmission deadline, and queue stability etc, but is not considered in this work to maintain tractability. Readers are referred to \cite{Yang2010,Ozel2011,Sharma2010} for further studies on this topic.}
%In general, the optimal energy-aware transmission policy may involve continuous power control in accordance with the data traffic, transmission deadline, and queue stability etc, which is beyond the scope of this work.\cite{Yang2010,Ozel2011,Sharma2010}
The efficacy of the proposed schemes are demonstrated by numerical simulations.

In the past, sensor deployment problems have been examined mostly for sensors without energy harvesting capabilities (see \cite{Younis2008} for a survey on these topics). In particular, without energy harvesting considerations, sensor deployment policies have been proposed with the goal of minimizing the field reconstruction error in \cite{Yang2008}, of guaranteeing connectivity in \cite{ISHIZUKA2004}, and of maximizing sensor coverage in \cite{Huang2003}. In these works, sensors were assumed to be placed at deterministic locations, in which case, the task of finding the optimal sensor placement is often NP hard \cite{Younis2008}. Therefore, heuristic or approximate solutions were proposed to reduce the computational complexity. Due to advances in energy harvesting technology \cite{Raghunathan2006, Seah2009, Sudevalayam2011}, similar problems have also been examined recently, e.g., in \cite{ZhiAngEu2009} and \cite{Misra2011}, for energy harvesting wireless systems with considerations on the stochastic nature of the energy arrival at each node. In particular, in \cite{ZhiAngEu2009} and \cite{Misra2011}, the deployment of energy harvesting relay nodes in sensor networks were examined with the goal of enhancing network throughput and of guaranteeing connectivity, respectively. However, they did not consider the cross-layer impact of the spatial dependencies of the energy arrivals and sensor measurements on the field reconstruction performance. In our work, we consider the optimal random sensor {\black deployment} strategy for field reconstruction by taking into consideration the spatial correlation between the sensor measurements and the energy arrivals. Rather than designing a deterministic deployment scheme which places sensors at precise locations, we consider the random deployment of sensors and determine the optimal sensor {\black densities} at different locations. We argue that random deployment \cite{Kwon2013, Tang2006} is more practical when the number of sensors to be deployed is large or when the sensors are to be deployed in a hostile environment. Comparisons between deterministic and random deployment schemes can also be found in \cite{Zhang2006} and \cite{Balister2009}.
%{\black Finally, it is noted that this work focuses on the deployment of sensors given the fixed and known data-gathering node locations, because the joint optimization of the data-gathering node locations and sensor deployment densities is extremely complicated. For those interested in the deployment of data-gathering nodes in sensor networks without energy harvesting capability, please see \cite{Kwon2013}.}

Field reconstruction and decentralized parameter estimation are essential applications of WSNs and have been studied extensively in the literature for WSNs without energy harvesting capabilities, e.g., in \cite{Bahceci2008, Fang2009, Behbahani2012, Chaudhary2012, Nevat2013}. For given sensor locations, these works focused on the design of sensor transmission schemes under different centralized fusion rules. Two transmission systems have been considered the most in the literature, namely, AF \cite{Bahceci2008, Fang2009, Behbahani2012} and DF {\black{\cite{Chaudhary2012,Nevat2013,Matamoros2013}}} systems. {\black In these systems,
%sensors simply transmit amplified versions of their noisy local observations to the data-gathering node(s). In this case,
the amplifying gains \cite{Bahceci2008, Fang2009, Behbahani2012} and the number of quantization bits \cite{Chaudhary2012, Nevat2013,Matamoros2013} used for transmission by the sensors respectively can be chosen to minimize the field reconstruction error.}
%at the sensors  \cite{Bahceci2008, Fang2009, Behbahani2012}; whereas, in DF systems, sensor measurements are quantized and encoded using numbers of bits that depend on the energy and bandwidth limitations of the system {\black\cite{Chaudhary2012, Nevat2013,Matamoros2013}}.
Both of AF and DF systems are examined in our work under additional energy harvesting considerations. Decentralized estimation under energy harvesting constraints have also been investigated recently in \cite{Liu2014} and \cite{Nayyar2013}. However, the sensor {\black deployment} problem and the impact of the location-dependent sensor measurements and energy arrivals on the estimation performance have not been explored before. A preliminary version of our work can be found in \cite{Hsu2013}, but only for the AF case.

The rest of this paper is organized as follows. A general description of the system model is presented in Section \ref{sec:system}. The sensor deployment and transmission control policies are then derived separately for AF and DF systems in Sections \ref{sec:Opt_Deploy_and_TC_AF} and \ref{sec:Opt_Deploy_and_TC_DF}, respectively. Numerical results are provided in Section \ref{sec:sim} and some concluding remarks are given in Section \ref{sec:conc}.

%%%%%%%%%%%%%%%%%%%%%%%%%%%%%%%%%%%%%%%%%
%% System Model and Problem Definition  %
%%%%%%%%%%%%%%%%%%%%%%%%%%%%%%%%%%%%%%%%%
\section{System Model and Problem Definition}\label{sec:system}
\begin{figure}[t]
     \centering
     \includegraphics[scale=0.42]{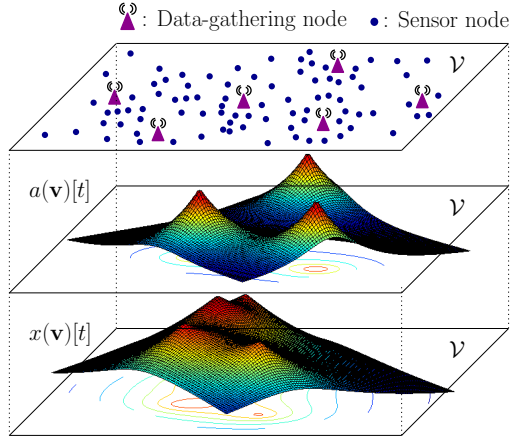}
     \caption{Illustration of the sensor deployment, the energy arrival distribution and the field values for a specific $t$.}
     \label{fig:deployment_field_reconstruction}
    \vspace{-.3cm}
\end{figure}

Let us consider a WSN that consists of a large number of sensors deployed randomly in the region $\calV$ according to a spatially NHPPP \cite{Streit2010} with deterministic {\black densities} $\{\lambda(\bv)\}_{\bv\in\calV}$, where $\lambda(\bv)$ is the {\black density} at location $\bv\in\calV$. In an observation period, each sensor takes a local sample of the random field and forwards it to the closest data-gathering node. {\black The locations of data-gathering nodes can be arbitrary but are assumed to be fixed over time.} The random field is denoted by $\{x(\mathbf v)[t]\}_{\mathbf v\in\mathcal V}$, where $x(\bv)[t]$ is the value of the field at location $\bv\in\calV$ in the $t$-th observation period. Following \cite{Yang2008} and \cite{Nevat2013}, we assume that the set of random field values $\{x(\mathbf v)[t]\}_{\mathbf v\in\mathcal V}$ is jointly Gaussian with zero mean and with covariances $E[x(\mathbf v_i)[t]x(\mathbf v_j)[t]]=\sigma_{x}^2\rho(\bv_i,\bv_j)[t]$, for all $\mathbf v_i,\mathbf v_j\in\mathcal V$, where $\sigma_x^2$ is the variance of $x(\mathbf v)[t]$, for all $\bv\in\calV$, and $\rho(\bv_i,\bv_j)[t]$ is the correlation coefficient between $x(\mathbf v_i)[t]$ and $x(\mathbf v_j)[t]$. {\black The random field is also assumed to be stationary over time.} Moreover, let $\{a(\bv)[t]\}_{\bv\in\calV}$ be the set of random energy arrivals, where $a(\bv)[t]$ is the energy arrival at location $\bv\in\calV$ in the $t$-th observation period. An example of the sensor deployment, the field values, and the energy arrivals for a specific $t$ are depicted in Fig. \ref{fig:deployment_field_reconstruction}. In practice, the set of sensor {\black densities} $\{\lambda(\bv)\}_{\bv\in\calV}$ to be derived in this work provides a guideline for the numbers of sensors that should be scattered at different locations, e.g., from an air vehicle.

To simplify our computations, let us partition the region of interest, i.e., $\mathcal V$, into $M$ disjoint subregions $\mathcal V_1,\ldots,\mathcal V_M$ with equal size. That is, we choose $\mathcal V_1,\ldots,\mathcal V_M$ such that $\mathcal V=\bigcup_{i=1}^M\mathcal V_i$ with areas $|\mathcal V_i|=\Delta$, for all $i$, and with $\mathcal V_i\bigcap\mathcal V_j=\phi$, for all $i\neq j$. We assume that $\Delta$ is sufficiently small (compared to the variations of the random field) so that the sensor {\black densities} remain approximately constant in each subregion (\i.e., $\lambda(\bv)\approx\lambda_i$, for all $\bv\in\calV_i$ and for all $i$). Then, by assuming that $\lambda_i\Delta\ll 1$, the probability that exactly one sensor exists in subregion $\calV_i$ can be approximated as $\frac{e^{-\lambda_i\Delta}\lambda_i\Delta}{1!}\approx\lambda_i\Delta$ and the probability that more than one sensor exists in $\calV_i$ is negligible. Moreover, with $\Delta$ sufficiently small, we can also approximate the field value and the energy arrival at a sensor in subregion $\calV_i$  by their respective values at the center of the subregion, which are denoted by $x_i[t]$ and $a_i[t]$, respectively. Therefore, if a sensor exists in subregion $\calV_i$, its local observation can be expressed as
\begin{equation}
\tilde{x}_i[t]=x_i[t]+n_{i}[t],
\end{equation}
where $n_i[t]$ is the observation noise at the sensor located at the subregion $\calV_i$. The observation noise is assumed to be independent and identically distributed (i.i.d.) across sensors {\black and over time} according to a Gaussian distribution with mean zero and variance $\sigma_n^2$, denoted by $\calN(0,\sigma_n^2)$.

To preserve energy at the sensors, we adopt a threshold-based energy-aware transmission policy where a sensor transmits only if its accumulated energy exceeds a certain threshold, but exhausts all its energy when doing so. This scheme is referred to in the literature as the \emph{integrate-and-fire} \cite{Scaglione2003,Huang2013} transmission policy. The energy threshold at the sensor in subregion $\calV_i$ is denoted by $\gamma_i$ and is chosen to minimize the field reconstruction error in later sections. The accumulated energy at each sensor, say, the sensor in subregion $\calV_i$, varies over time $t$ and can be expressed as
\begin{equation}\label{eq:ei}
e_i[t]=\left\{\begin{array}{cc}e_i[t-1]+a_i[t], &\text{ if } e_i[t-1]\leq \gamma_i\\
a_i[t], &\text{ if } e_i[t-1]> \gamma_i.\end{array}\right.
\end{equation}
at the beginning of the $t$-th observation period. By assuming that the process $\{e_i[t]\}_{t=0}^\infty$ is stationary over time, the probability that the sensor transmits in a given time slot is $\bar{F}_{e_i}(\gamma_i)\triangleq\Pr(e_i[t]>\gamma_i)$, which is the complementary cumulative density function (ccdf) of $e_i[t]$. {\black An example of the case where $\{e_i[t]\}_{t=0}^\infty$ is stationary is given in the following sections by considering a Bernoulli energy arrival process.} By the stationarity of the random field $\{x_i[t]\}_{t=0}^\infty$ and of the accumulated energy process $\{e_i[t]\}_{t=0}^\infty$, for all $i$, we shall omit the time index $t$ in later discussions and focus on the field reconstruction performance at a particular instant in time.
%{\black This paper also assumes that the
%random field $\{x(\bv)[t]\}_{t=0}^\infty$ for $\bv\in\calV$ is stationary over time. As a result, we shall omit the index $t$ in later discussions, because the obtained result is the same for all time $t$ due to the stationarity of the random field and the accumulated energy.}

Let us define a random variable $o_i$ to indicate the presence or absence of a signal transmitted by a sensor in subregion $\calV_i$. Specifically, we set $o_i=1$, if a sensor exists in subregion $\calV_i$ and has energy above the threshold $\gamma_i$, and set $o_i=0$, otherwise. In this case, $o_i$ can be viewed as a Bernoulli random variable with probability $\Pr(o_i=1)=\lambda_i\Delta\bar{F}_{e_i}(\gamma_i)$ and $\Pr(o_i=0)=1-\lambda_i\Delta\bar{F}_{e_i}(\gamma_i)$. Moreover, let $r_i$ be the signal received from the sensor in subregion $\calV_i$ at the closest data-gathering node over a noisy channel. When $o_i=1$, the received signal $r_i$ depends on the the sensor's observation $\tilde{x}_i$, the accumulated energy $e_i$, the type of transmission scheme, and the quality of the channel towards the closest data-gathering node. When $o_i=0$, nothing is received {\black and, thus, we set $r_i=\text{null}$}. Two types of sensor transmission schemes are considered, namely, the AF and DF schemes. In the AF scheme, each sensor simply transmits an amplified version of its received signal to the closest data-gathering node. In the DF scheme, each sensor first quantizes its measurement into a binary representation vector and then forwards it to the closest data-gathering node, where the binary vector is decoded and the quantized sensor measurement is reconstructed.

The signals received at the data-gathering nodes are transmitted over the backhaul to the fusion center where an estimate of the field values $\{x(\bv)\}_{\bv\in\calV}$ are computed. Let $\bo\triangleq [o_1,\ldots, o_M]^T$ be the vector representing the sensors' transmission status. The transmissions from the data-gathering nodes to the fusion center are assumed to be error-free and, thus, $\{r_i\}_{i=1}^M$ are also the observations available at the fusion center. {\black By assuming that the fusion center is aware of the sensors' locations and transmission status\footnote{\black Here, we assume that the fusion center is aware of the locations of the sensors once they have been deployed and the transmission status of the existing sensors. The former can be obtained through positioning techniques after deployment and the latter can be obtained by appending the sensors' IDs to their transmissions or by performing signal detection at the fusion center.
%(e.g., if the fusion center is not able to synchronize with any signal transmission, then it will infer that no sensor is transmitting).
However, the proposed methodology can also be used for the case where no knowledge is assumed at the fusion center. In this case, the received signal in the absence of a sensor transmission in subregion ${\cal V}_i$ should be $r_i=w_i$.}, i.e., $\{o_i\}_{i=1}^M$,
it extracts only the set of field-bearing observations $\{r_i|1\leq i\leq M, o_i=1\}$ for field reconstruction. The linear minimum mean square error (LMMSE) estimator is then adopted to obtain an estimate of each point in the field using these observations. Interestingly, we show in Appendix \ref{APP:equiv} that the LMMSE estimate obtained with the set of observations $\{r_i|1\leq i\leq M, o_i=1\}$ is equivalent to that obtained with the effective received signal vector $\by\triangleq [y_1,\ldots, y_M]^T$, where
\[
y_i=\left\{\begin{array}{ll}r_i, &\text{ if } o_i=1\\
0, &\text{ if } o_i=0,\end{array}\right.
\]
for all $i$. Therefore, the estimate of the field value $x(\bv)$ at location $\bv$ can be written as
%. In particular, by letting $\by\triangleq [y_1,\ldots, y_M]^T$ be the effective received signal vector, the LMMSE estimate can be equivalently written as
%The resulting linear minimum mean square error (LMMSE) estimate can be equivalently written as
%}{\black It is assumed that the fusion center has the knowledge of the sensor's transmission status, i.e., $\{o_i\}_{i=1}^M$,
%when performing the field reconstruction.
%\footnote{\black In practice, the knowledge of sensor's transmission status can be extracted, for example, by TDMA, when the received signal-to-noise ratio (SNR) is high enough to distinguish between the noisy signal and the pure noise.} The distortion of field reconstruction is quantified by the mean-square error (MSE) defined as $E\left[|x(\bv)-\hat x(\bv)|^2\left|\bo\right.\right]$, where $\hat x(\bv)$ is the estimate of the field value $x(\bv)$ at location $\bv\in\calV$. Note that, for an accurate field reconstruction, it is reasonable not to use the noise-only observations, i.e., $\{r_i|1\leq i\leq M, o_i=0\}$. Consequently, for convenience, we define $\by\triangleq [y_1,\ldots, y_M]^T$, where
%\begin{equation*}
%y_i=\left\{\begin{array}{ll}r_i, &\text{ if } o_i=1\\
%0, &\text{ if } o_i=0.\end{array}\right.
%\end{equation*}
%The linear minimum mean-square error (LMMSE) estimator $\hat x(\bv)$ using only the field-value-bearing received signal $\by$ can be equivalently written as}
\begin{align}
\hat{x}(\bv)=\mathbf C^\bo_{x(\bv)\mathbf y}{\bC_{\by\by}^\bo}^{\!\!\dag}\mathbf y, \label{eq:LMMSE_estimator}
\end{align}
}where
%$\by\triangleq [y_1,\ldots, y_M]^T$ with
%\[
%y_i=\left\{\begin{array}{ll}r_i, &\text{ if } o_i=1\\
%0, &\text{ if } o_i=0,\end{array}\right.
%\]
%for all $i$, can be viewed as the effective received signal vector,}
$\bC_{\by\by}^\bo\triangleq E[\by\by^T|\bo]$, $\mathbf C^\bo_{x(\bv)\mathbf y}\triangleq E[x(\bv)\by^T|\bo]$, and $^\dag$ represents the Moore-Penrose pseudo inverse.
%{\red The proof of the equivalence between the expression of the LMMSE estimate in %{\black (The proof of equivalence of the LMMSE in
%\eqref{eq:LMMSE_estimator} and that obtained with the observation set $\{r_i|1\leq i\leq M, o_i=1\}$ is given in Appendix \ref{APP:equiv}.}
The resulting MSE of the estimate on $x(\bv)$ is then given by
\begin{align}
{\black\xi(\mathbf v|\bo)%&=\min_{\bk(\bv)\in\mathbb{R}^M}E\left[\|x(\bv)-\bk(\bv)^T\by\|^2\left|\bo\right.\right]\\
=\sigma_x^2-\tr\left(\bC^\bo_{x(\bv)\mathbf y}{\bC_{\by\by}^\bo}^{\!\!\dag}{\bC^\bo_{x(\bv)\mathbf y}}^{\!\!\!\!T}\right)}
\end{align}
and the
%The
average MSE over the entire sensor field is defined as
%, defined as
%the average of the {\black MSE} over all points of the sensor field, is
%then given by
\begin{equation}
{\black\bar\xi\triangleq\frac{1}{|\mathcal V|}\int_{\mathbf v\in\mathcal V}\sum_{\bo\in\calO}\xi(\bv|\bo)\Pr(\bo)d\bv,\label{eq:avMSE}}
\end{equation}
where $\calO$ is the set of all possible realizations of $\bo$. Notice that the above expression involves the summation over all possible realizations of $\bo$, which can be intractable in practice. To simplify our computations, we consider instead an upper bound of the average MSE given as follows:
{\black\begin{align}
\bar\xi&{\black=\frac{1}{|\calV|}\int_{\bv\in\calV}E_\bo\left[\min_{\bk\in\mathbb{R}^M}E\left[|x(\bv)-{\bk}^T\by|^2\left|\bo\right.\right]\right]d\bv\label{eq:avMSE1}}\\
&\leq\frac{1}{|\calV|}\int_{\bv\in\calV}\min_{\bk\in\mathbb{R}^M}E_\bo\left[E\left[|x(\bv)-{\bk}^T\by|^2\left|\bo\right.\right]\right]d\bv\label{eq:avMSEupper_min_given_o}\\
&=\frac{1}{|\mathcal V|}\int_{\mathbf v\in\mathcal V}\min_{{\bk}\in\mathbb{R}^M}E\left[{\black|}x(\bv)-{\bk}^T\by{\black|}^2\right]d\bv\label{eq:avMSEupper_min}\\
&=\frac{1}{|\mathcal V|}\int_{\mathbf v\in\mathcal V}\!\!\sigma_x^2-\tr\left(\!\bC_{x(\bv)\mathbf y}\bC_{\by\by}^{-1}\bC_{x(\bv)\mathbf y}^T\!\right)d\bv\triangleq \bar\xi_{\rm upper},\label{eq:avMSEupper_general}
\end{align}
where $\bC_{x(\bv)\by}\triangleq E[x(\bv)\by^T]$ and $\bC_{\by\by}\triangleq E[\by\by^T]$ are computed by taking the expectation over all possible realizations of $\bo$, and thus, do not depend on the actual subset of transmitting sensors. The vector $\bk$ inside the expectation in \eqref{eq:avMSE1} can be viewed as the linear estimator that should be optimized separately for each given value of $\bv$ and $\bo$. The inequality in \eqref{eq:avMSEupper_min_given_o} follows since the minimization over each term inside the expectation must be smaller than the minimization over the entire expectation (which amounts to using the same $\bk$ for each $\bo$). The latter can be interpreted as the minimum MSE attainable when the fusion center does not have knowledge of the sensors' locations and transmission status. This bound is tight only when the optimal linear estimator $\bk$ is approximately the same for all realizations of $\bo$  that occur with high probability.
%where {\black the inequality in \eqref{eq:avMSEupper_min_given_o} follows from {\black moving the minimization outside the expectation}.
Even though the sensors are deployed randomly, their locations are fixed once they are deployed and do not change over time. The MSE in \eqref{eq:avMSE1} provides a measure of the average performance over all possible sets of sensor locations following an NHPPP with sensor densities $\{\lambda_i\}_{i=1}^M$, but is only an approximation of the actual MSE in practice, which corresponds to just one realization of $\bo$.}

%In practice, sensors are deployed only once according to the sensor densities $\lambda
%{\black Here, $\bC_{x(\bv)\by}\triangleq E[x(\bv)\by^T]$ and $\bC_{\by\by}\triangleq E[\by\by^T]$ are computed by taking the expectation over all the possible realizations of $\{o_i\}_{i=1}^M $, and thus, do not depend on the actual subset of transmitting sensors. We would like to note that, in practice, sensors are placed and fixed to perform field reconstruction. Hence, the average MSE in \eqref{eq:avMSE} provides only an approximation to the actual MSE in the case where sensors are placed once and then stay at the fixed locations. However, the average MSE provides a guideline for determining the optimal sensor densities in a sensing field. Therefore, t}

The main objective of this work is to determine the optimal sensor {\black densities} $\{\lambda_i\}_{i=1}^M$ and the energy thresholds $\{\gamma_i\}_{i=1}^M$ by minimizing the MSE upper bound in \eqref{eq:avMSEupper_general} for both the AF and the DF systems. The optimized sensor {\black densities} should achieve the optimal balance between the energy arrival probability, the channel gain, and the sensor field correlation whereas the optimized energy thresholds should exploit the tradeoff between the sensor's transmission probability and the reception quality at the data-gathering nodes. The two systems are examined separately in the following sections.

%%%%%%%%%%%%%%%%%%%%%%%%%%%%%%%%%%%%%%%%%%%%%%%%%%%%%%%%%%%%%%%%%%%%%%%%%%%%%%%%%%%%%%%%%%%%%%%
%% Optimized Sensor Deployment and Transmission Control Strategy for Analog Forwarding System %
%%%%%%%%%%%%%%%%%%%%%%%%%%%%%%%%%%%%%%%%%%%%%%%%%%%%%%%%%%%%%%%%%%%%%%%%%%%%%%%%%%%%%%%%%%%%%%%
\section{Optimized Sensor {\black Densities} and Energy Thresholds for Analog-Forwarding Systems}\label{sec:Opt_Deploy_and_TC_AF}

In this section, we determine the optimal sensor {\black densities} $\{\lambda_i\}_{i=1}^M$ and energy thresholds $\{\gamma_i\}_{i=1}^M$ in AF systems based on the minimization of the average MSE upper bound in \eqref{eq:avMSEupper_general}.

\begin{figure}[t]
     \centering
     \includegraphics[scale=0.6]{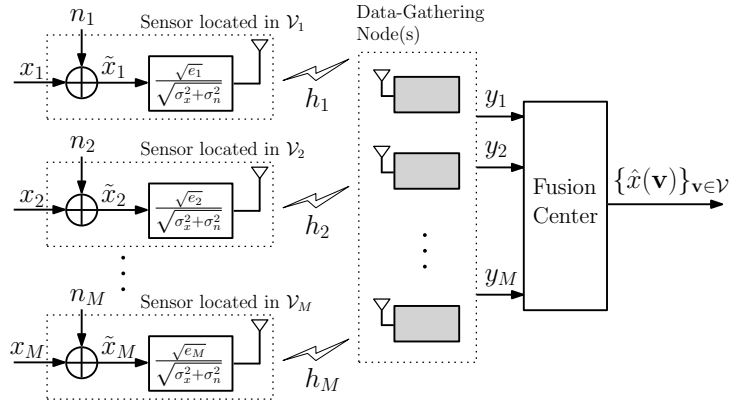}
     \caption{Illustration of filed reconstruction for the analog forwarding system.}
     \label{fig:system_model_AF}
     \vspace{-.3cm}
\end{figure}

In the AF system, each sensor transmits an amplified version of its local observation to the closest data-gathering node, as illustrated in Fig. \ref{fig:system_model_AF}.  In this case, the signal transmitted by a sensor in subregion $\calV_i$ can be written as
\begin{align}
{\black s_i=\frac{\sqrt{\kappa_i e_i}}{\sqrt{\sigma_x^2+\sigma_n^2}}\, \tilde{x}_i,\label{eq:si}}
\end{align}
where {\black $\kappa_i>0$ is the amplifying factor}, $e_i$ is the energy available at the sensor in $\calV_i$, and the received signal at the closest data-gathering node is
\begin{align}
{\black y_i=\left(h_i\frac{\sqrt{\kappa_ie_i}}{\sqrt{\sigma_x^2+\sigma_n^2}}\,\tilde{x}_i+w_i\right)o_i,}
\end{align}
where $h_i$ is the channel coefficient between the sensor and its closest data-gathering node{\black\footnote{\black The signal model under consideration can also accommodate broadcast transmissions, where the signal transmitted by a sensor in subregion $\calV_i$ may be received simultaneously by multiple data-gathering nodes simultaneously. In this case, $h_i$ can be viewed as the effective channel coefficient after signal combining at the fusion center.}}, and $w_i\sim\calN(0,\sigma_w^2)$ is the additive white Gaussian noise (AWGN). Here, we do not consider the effects of short-term fading and assume that $h_i$ is a constant that depends only on the distance between the sensor and its data-gathering node. Therefore, by defining $g_i\triangleq\frac{h_i\sqrt{\kappa_ie_i}}{\sqrt{\sigma_x^2+\sigma_n^2}}$ as the effective channel gain experienced by the sensor in $\calV_i$, the received signal is expressed as $y_i=(g_i\tilde x_i+w_i)o_i$. {\black For simplicity, we assume that the sensor depletes its battery every time it transmits (i.e., any residual energy is dumped after each transmission). Notice that, since the observation $\tilde x_i$ is random, the energy required to transmit the signal in \eqref{eq:si} may exceed the accumulated energy $e_i$ with a certain probability. In this case, a saturation effect may occur, causing additional distortion on the received signal. However, by choosing $\kappa_i$ to be sufficiently small, the probability that the saturation occurs is small and, thus, is omitted for simplicity. Studies on the impact of this effect on the distributed estimation performance is beyond the scope of this work, but can be found in \cite{Hong2014}.}
%{\black Notice that, the observation of sensor may not be {\black forwarded to the data-gathering nodes with the distortionless value of \eqref{eq:si}} when the instantaneous power of the noisy observation $\tilde x_i$ is large such that $\frac{\kappa_i\tilde x_i^2}{\sigma_x^2+\sigma_n^2}>1$. However, we can properly chose $\kappa_i$ to make the probability $\Pr\left(\frac{\kappa_i\tilde x_i^2}{\sigma_x^2+\sigma_n^2}>1\right)=\calQ\left(\sqrt{\frac{1}{\kappa_i}}\right)$ sufficiently small to neglect this effect\footnote{{\black The sophisticated model to fully characterize the effect of the amplifying factor $\kappa_i$ leads complicated formulation and is difficult to handle in our sensor deployment problem. The incorporation of $\kappa_i$ in distributed estimation in energy harvesting wireless sensor network is discussed in our previous work \cite{Hong2014}.}}.
%In this case, by} defining $g_i\triangleq\frac{h_i\sqrt{{\black\kappa_i}e_i}}{\sqrt{\sigma_x^2+\sigma_n^2}}$ as the effective channel gain experienced by the sensor in $\calV_i$, the received signal is expressed as $y_i=(g_i\tilde x_i+w_i)o_i$.

For convenience, let $\alpha_i\triangleq \Pr(o_i=1)=\bar{F}_{e_i}(\gamma_i)\lambda_i\Delta$ be the probability that a sensor exists in $\calV_i$ and transmits in the given observation period. Moreover, let $\bar g_i(\gamma_i)\triangleq E[g_i(\gamma_i)|e_i>\gamma_i]$ and $\overline{g_i^2}(\gamma_i)\triangleq E[g_i^2(\gamma_i)|e_i>\gamma_i]$ be the conditional first and second moments of the effective channel gain given that the sensor transmits. In this case, we have $\bC_{x(\bv)\by}=\bC_{x(\bv)\bx}\bD_{\bar g}\bD_\alpha$, where $\bD_{\bar g}\triangleq\mbox{diag}\left(\bar g_1(\gamma_1),\ldots, \bar g_M(\gamma_M)\right)$ and $\bD_\alpha=\mbox{diag}(\alpha_1,\ldots,\alpha_M)$; thus, the MSE upper bound in \eqref{eq:avMSEupper_general} can be written as
\begin{equation}
\bar\xi_{\rm upper, AF}=\sigma_x^2-\tr\left(\bPhi\bD_{\bar g}\bD_\alpha\bC_{\by\by}^{-1}\bD_\alpha\bD_{\bar g}\right)\label{eq:avMSEupper},
\end{equation}
where $\bPhi\triangleq\frac{1}{|\mathcal V|}\int_{\mathbf v\in\mathcal V}\bC_{x(\bv)\bx}^T \bC_{x(\bv)\bx}d\bv$ is defined such that
\begin{equation}
\{\bPhi\}_{i,j}=\frac{\sigma_x^4}{|\mathcal V|}\int_{\mathbf v\in\mathcal V}\rho(\bv,\bv_i)\rho(\bv,\bv_j)d\bv\triangleq \phi_{i,j},\label{eq:Phi_ij}
\end{equation}
for all $i,j$, and the $(i,j)$-th element of $\bC_{\by\by}$ can be derived as
\begin{equation}
\{\!\bC_{\by\by}\!\}_{i,j}\!=\!\left\{\!\!\!\begin{array}{cc}[\overline{g_i^2}(\gamma_i)(\sigma_x^2+\sigma_n^2)+\sigma_w^2]\alpha_i, & \text{for } i\!=\!j\\
\bar g_i(\gamma_i)\bar g_j(\gamma_j)\sigma_x^2\rho(\bv_i,\!\bv_j)\alpha_i\alpha_j, & \text{for } i\!\neq\! j,\end{array}\right.
\end{equation}
for all $i,j$. Notice that $\Phi$ is a matrix that depends only on the correlation of the sensor field and not on the optimizing parameters $\{\lambda_i\}_{i=1}^M$ and $\{\gamma_i\}_{i=1}^M$. Let us take the eigenvalue decomposition of $\bPhi$ so that $\bPhi=\mathbf U\boldsymbol\Sigma\mathbf U^T$, where $\boldsymbol\Sigma$ is a diagonal matrix consisting of the eigenvalues of $\bPhi$ and $\mathbf U$ is a unitary matrix consisting of the corresponding eigenvectors. In this case, the term in \eqref{eq:avMSEupper} can be written as
\begin{equation}
\bar\xi_{\rm upper, AF}=\sigma_x^2-\tr\left(\boldsymbol\Sigma^\frac{1}{2}\bU^T\bD_{\bar g}\bD_\alpha\bC_{\by\by}^{-1}\bD_\alpha\bD_{\bar g}\bU\boldsymbol\Sigma^\frac{1}{2}\right).\label{eq:avMSEupper_AF_eigen}
\end{equation}

To further facilitate our derivations, we introduce the following lemma from \cite{Fang2009}, which is a consequence of the Cauchy-Schwarz inequality.
\begin{lemma}[\!\!\cite{Fang2009}]\label{lemma:Cauchy_Schwarz}
For any $\bG\in \mathbb R^{M\times K}$ and positive-definite matrix $\mathbf Q\in R^{M\times M}$, the following inequality holds:
\begin{align}
\tr(\mathbf G^T\mathbf Q^{-1}\mathbf G)\geq\frac{\left[\tr(\mathbf G^T\mathbf G)\right]^2}{\tr(\mathbf G^T\mathbf Q\mathbf G)}.
\end{align}
{\black The equality holds when $\bG^T\bQ^{-\frac{1}{2}}=c\bG^T\bQ^{\frac{1}{2}}$, where $c$ is a constant.}
\end{lemma}
By taking $\mathbf G=\bD_\alpha\bD_{\bar g}\bU\boldsymbol\Sigma^{\frac{1}{2}}$ and $\mathbf Q=\bC_{\by\by}$, it follows that
\begin{align}
\bar\xi_{\rm upper, AF}
&\leq\sigma_x^2\!-\!\frac{\left[\tr\left(\boldsymbol\Sigma^{\frac{1}{2}}\bU^T(\bD_{\bar g}\bD_\alpha)^2\bU\boldsymbol\Sigma^{\frac{1}{2}}\right)\right]^2}{\tr\left(\boldsymbol\Sigma^{\frac{1}{2}}\bU^T\bD_{\bar g}\bD_\alpha\bC_{\by\by} \bD_\alpha\bD_{\bar g}\bU\boldsymbol\Sigma^{\frac{1}{2}}\right)}\label{eq:avMSEupper2pre}\\
&=\sigma_x^2-\frac{\left[\tr\left(\bPhi(\bD_{\bar g}\bD_\alpha)^2\right)\right]^2}{\tr\left(\bPhi\bD_{\bar g}\bD_\alpha\bC_{\by\by}\bD_\alpha\bD_{\bar g}\right)}\triangleq \bar\xi_{\rm obj, AF}(\boldsymbol\lambda,\boldsymbol\gamma),\label{eq:avMSEupper2}
\end{align}
where $\boldsymbol\lambda=[\lambda_1,\ldots,\lambda_M]^T$ is the vector of sensor {\black densities} and $\boldsymbol\gamma=[\gamma_1,\ldots,\gamma_M]^T$ is the vector of energy thresholds.
{\black By Lemma \ref{lemma:Cauchy_Schwarz}, the bound is tight when $(\bD_\alpha\bD_{\bar g}\bU\boldsymbol\Sigma^{\frac{1}{2}})^T\bC_{\by\by}^{-(1/2)}=c(\bD_\alpha\bD_{\bar g}\bU\boldsymbol\Sigma^{\frac{1}{2}})^T\bC_{\by\by}^{1/2}$, for some constant $c$. The upper bound $\bar\xi_{\rm obj, AF}(\boldsymbol\lambda,\boldsymbol\gamma)$ is then utilized as the objective function for optimizing $\boldsymbol\lambda$ and $\boldsymbol\gamma$ in the AF case. Even though the bound may not be tight in general, it captures the essential behaviors of the MSE with respect to the sensor densities and transmission thresholds and, thus, allow us to obtain a tractable solution to an otherwise intractable problem. The effectiveness of the solution is demonstrated in Section \ref{sec:sim}.}

%is a function of $\boldsymbol\lambda$ and $\boldsymbol\gamma$, where $\boldsymbol\lambda=[\lambda_1,\ldots,\lambda_M]^T$ is the vector of sensor {\black densities} and $\boldsymbol\gamma=[\gamma_1,\ldots,\gamma_M]^T$ is the vector of energy thresholds. This upper bound is then utilized as the objective function for optimizing $\boldsymbol\lambda$ and $\boldsymbol\gamma$ in the AF case.

Specifically, let us consider the optimization problem where the MSE upper bound
%Our problem is then to minimize the MSE upper bound
$\bar\xi_{\rm obj, AF}$ in \eqref{eq:avMSEupper2} is minimized subject to a constraint on the average total number of sensors, i.e.,
\begin{subequations}\label{eq:problem_formulation}
\begin{align}
\min_{\boldsymbol\lambda,\boldsymbol\gamma} \quad&\bar\xi_{\rm obj, AF}(\boldsymbol\lambda,\boldsymbol\gamma)\\
\mbox{subject to}\quad&\sum_{i=1}^M\lambda_i\Delta\leq\bar\Lambda,\\
&0<\lambda_i\Delta{\black\leq\epsilon_{\Lambda}},~\text{for }i=1,\ldots, M.
\end{align}
\end{subequations}
where $\bar\Lambda$ is the constraint on the average number of sensors in the network. Here, the approximate probability that a sensor exists in subregion $\calV_i$, i.e., $\lambda_i\Delta$, is restricted within {\black$[0,\epsilon_\Lambda]$} and {\black $\epsilon_\Lambda<1$ should be small enough to make the approximation accurate enough.} The value of $\lambda_i$ is assumed to be positive to ensure that $\bC_{\by\by}$ is invertible.
{\black Following the relation in \eqref{eq:ei}, the energy threshold $\boldsymbol\gamma$ determines the distribution of the accumulated energy process and, thus, affects the objective function through the probabilities $\alpha_i$, for all $i$, which involve the event that a sensor exists and transmits in each subregion.}

Since the first term in \eqref{eq:avMSEupper2} is a constant, the optimization problem can be equivalently formulated as the maximization over the second term in \eqref{eq:avMSEupper2}. By the change of variable $\Lambda_i=\lambda_i\Delta$ and by defining $\bLambda=[\Lambda_1,\ldots, \Lambda_M]^T$, the optimization problem can be written explicitly as follows:
\begin{subequations}\label{opt:max}
\begin{align}
\max_{\bLambda,\boldsymbol\gamma} \quad&J_{\rm AF}(\bLambda,\bgamma)={\frac{J^{\text{num}}_{\rm AF}(\bLambda,\bgamma)}{J^{\text{den}}_{\rm AF}(\bLambda,\bgamma)}}\\
\mbox{subject to}\quad&\sum_{i=1}^M\Lambda_i\leq\bar\Lambda\\
&0<\Lambda_i{\black\leq\epsilon_{\Lambda}},~\text{for }i=1,\ldots, M.
\end{align}
\end{subequations}
The term in the numerator of the objective function is
\begin{align}
J^{\text{num}}_{\rm AF}(\bLambda,\bgamma)=&\left(\sum_{i=1}^M\phi_{i,i}[\bar g_i(\gamma_i)]^2\bar F_{e_i}^2(\gamma_i)\Lambda_i^2\right)^2\label{eq:J_AF_Num}
\end{align}
and that in the denominator is
\begin{align}
\notag J^{\text{den}}_{\rm AF}(\bLambda,\bgamma)=&\sum\limits_{i=1}^M\!\phi_{i,i}\left[\overline{g_i^2}(\gamma_i)\sigma_{\tilde x}^2+\sigma_w^2\right]\bar{F}^3_{e_i}(\gamma_i)\Lambda_i^3\bar g_i^2(\gamma_i)\\
&\!\!+\!\!\sum_{i=1}^M\!\sum_{j\neq i}\!\!\phi_{i,j}\sigma_x^2\rho(\bv_i,\!\!\bv_j\!)\bar{F}^2_{e_i}\!(\gamma_i)\Lambda_i^2\bar g_i^2\!(\gamma_i)\bar{F}^2_{e_j}\!(\gamma_j)\Lambda_j^2\bar g_j^2\!(\gamma_j).\label{eq:J_AF_Den}
\end{align}
where $\sigma_{\tilde x}^2=\sigma_x^2+\sigma_n^2$.

Notice that the solution to the optimization problem in \eqref{opt:max} depends on the statistics of the energy arrival. In the following, we provide an example of how such a problem can be solved under the Bernoulli energy arrival model.

\subsection{Solution for Bernoulli Energy Arrival Case}\label{subsec:Solution_for_Bern_energy_arrival}

\begin{figure}[t]
     \centering
     \includegraphics[scale=0.55]{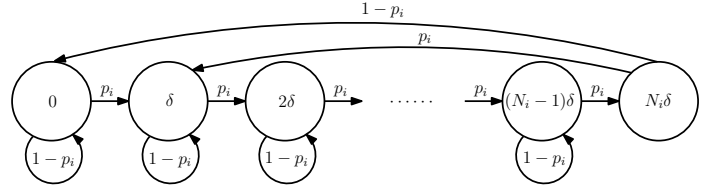}
     \caption{Energy harvesting model of Bernoulli energy arrival.}
     \label{fig:EH_Markov}
    \vspace{-.3cm}
\end{figure}

In this subsection, we apply the above techniques to cases where the energy arrival can be modeled as an i.i.d. Bernoulli process. {\black This model has been widely adopted in the literature, e.g., in \cite{Medepally2009,Michelusi2012,Kashef2012,Michelusi2013}, due to its tractability and because of its ability to model sources with sporadic energy arrivals in discrete-time. Examples of such energy sources may include vibrational, RF energy, or wind etc. It can also be used to approximate more stable energy sources, such as solar, by choosing each time slot to be sufficiently small so that energy will arrive approximately at a constant rate when viewed from a larger time-scale. Other energy arrival models can also fit into our framework, but may require different optimization techniques from that to be introduced in the following.}

%{\black A discrete-time Bernoulli process is a common model for describing energy arrival characteristic (see \cite{Medepally2009,Michelusi2012,Kashef2012,Michelusi2013}). It can also be used to capture the sporadic and random availability at an energy harvesting node. Therefore, we adopt the Bernoulli energy arrival process to obtain solutions for problem given in \eqref{opt:max}. Specifically, the energy arrival process $\{a_i[t]\}_{t=0}^\infty$ is assumed to be an i.i.d Bernoulli process with $\Pr(a_i[t]=\delta)=p_i$ and $\Pr(a_i[t]=0)=1-p_i,~\forall t$}. That is, the sensor in $\calV_i$ harvests energy $\delta$ in each observation period with probability $p_i$ and fails to harvest energy with probability $1-p_i$.

{\black Suppose that the energy arrival process $\{a_i[t]\}_{t=0}^\infty$ consists of a sequence of i.i.d. Bernoulli random variables with $\Pr(a_i[t]=\delta)=p_i$ and $\Pr(a_i[t]=0)=1-p_i,~\forall t$.} That is, the sensor in $\calV_i$ harvests energy $\delta$ in each observation period with probability $p_i$ and fails to harvest energy with probability $1-p_i$. Since the sensor expends all its energy once the energy threshold $\gamma_i$ is exceeded{\black\footnote{\black The accumulated energy may not be depleted after each transmission in the AF case since the required transmit power is random. However, we assume for simplicity that the remaining energy is omitted after each transmission.}}, {\black the accumulated energy $\{e_i[t]\}_{t=0}^\infty$ forms a $(N_i+1)$-state Markov process, as illustrated in Fig. \ref{fig:EH_Markov}, where $N_i\triangleq\min\{n:n\delta\geq \gamma_i\}$. The stationary (or steady-state) distribution is given by $\Pr(e_i[t]=0)=\frac{1-p_i}{N_i}$, $\Pr(e_i[t]=N_i\delta)=\frac{p_i}{N_i}$, and  $\Pr(e_i[t]=k\delta)=\frac{1}{N_i}$ for $k=1,\ldots,N_i-1$.} In this case, the probability that the sensor in subregion $\calV_i$ transmits (if it exists) is given by $\bar{F}_{e_i}(\gamma_i)=\Pr(e_i[t]=N_i\delta)$. Notice that, in the Bernoulli energy arrival model (c.f. Fig. \ref{fig:EH_Markov}), the accumulated energies are integer multiples of $\delta$ and, thus, the energy threshold $\gamma_i$ can also be set as a multiple of $\delta$, in which case, we have $\gamma_i\!=\!N_i\delta$. Consequently, we have $\bar{F}_{e_i}(\gamma_i)\!=\!\frac{p_i\delta}{\gamma_i}$, $\bar g_i(\gamma_i)\!=\!\frac{\tilde h_i\sqrt{\gamma_i}}{\sigma_{\tilde x}}$, and $\overline{g_i^2}(\gamma_i)\!=\!\frac{\tilde h_i^2\gamma_i}{\sigma_{\tilde x}^2}$, where $\tilde h_i\triangleq\sqrt{\kappa_i}h_i$. Notice that the stationary distribution as well as these values change as the energy threshold $\gamma_i$ is adjusted. By substituting the above into \eqref{eq:J_AF_Num} and \eqref{eq:J_AF_Den}, we get
\begin{align}
J_{\rm AF}^{\text{num}}(\bLambda,\bgamma)\!=\!\!(\sum_{i=1}^M\phi_{i,i} \tilde h_i^2 p_i^2\delta^2\Lambda_i^2\gamma_i^{-1})^2
\end{align}
and
\begin{align}
\notag
J_{\rm AF}^{\text{den}}(\bLambda,\bgamma)\!=&\!\sum_{i=1}^M\!\phi_{i,i}(\tilde h_i^2\gamma_i\!+\!\sigma_w^2)\tilde h_i^2\sigma_{\tilde x}^2p_i^3\delta^3\Lambda_i^3\gamma_i^{-2}\\
&
+\sum_{i=1}^M\!\sum_{j\neq i}\!\phi_{i,j}\sigma_x^2\rho(\bv_i,\!\bv_j)p_i^2\delta^2 \tilde h_i^2\Lambda_i^2\gamma_i^{-1}p_j^2\delta^2 \tilde h_j^2\Lambda_j^2\gamma_j^{-1}
\end{align}
and, thus, the optimization problem in \eqref{opt:max} can be written as
\begin{subequations}\label{opt:Bernoulli_AF}
\begin{align}
\max_{\boldsymbol\Lambda,\boldsymbol\gamma}~ &\frac{\displaystyle\sum_{i=1}^M\sum_{j=1}^M A_{i,j}\frac{\Lambda_i^2\Lambda_j^2}{\gamma_i\gamma_j}}{\displaystyle\sum_{i=1}^M\!C_i\frac{\Lambda_i^3}{\gamma_i}\!+\!\sum_{i=1}^M \!D_i\frac{\Lambda_i^3}{\gamma_i^2}\!+\!\!\sum_{i=1}^M\sum_{j\neq i}\!G_{i,j}\frac{\Lambda_i^2\Lambda_j^2}{\gamma_i\gamma_j}}\\
\mbox{subject to}&~\sum_{i=1}^M\Lambda_i\leq\bar\Lambda,\\
&~0<\Lambda_i{\black\leq\epsilon_{\Lambda}},\;i=1,\ldots,M,\\
&~\gamma_i=Z_i\delta,\; Z_i\in\mathbb{N},\;i=1,\ldots,M,
\end{align}
\end{subequations}
where $A_{i,j}\triangleq\phi_{i,i}\phi_{j,j}\tilde h_i^2\tilde h_j^2p_i^2p_j^2\delta^4~$, $C_i\triangleq\phi_{i,i}\tilde h_i^4p_i^3 \delta^3\sigma_{\tilde x}^2~$, $D_i\triangleq\phi_{i,i}\sigma_w^2p_i^3\delta^3\tilde h_i^2\sigma_{\tilde x}^2~$, and
$G_{i,j}\triangleq\phi_{i,j}\sigma_x^2\rho(\bv_i,\bv_j)p_i^2p_j^2\tilde h_i^2\tilde h_j^2\delta^4$.

Notice that this problem is a mixed integer nonlinear programming (MINLP) problem, which is difficult to solve in general. To address this issue, we consider a relaxation where the integer constraint on $\gamma_i$ is replaced with the inequality constraint $\gamma_i\geq \delta$. Then, by defining $f(\boldsymbol\Lambda,\boldsymbol\gamma)\triangleq\sum_{i=1}^M\sum_{j=1}^M A_{i,j}\frac{\Lambda_i^2\Lambda_j^2}{\gamma_i\gamma_j}$, the optimization problem can be written equivalently as
\begin{subequations}\label{opt:Bernoulli_AF_relax}
\begin{align}
\min_{\boldsymbol\Lambda,\boldsymbol\gamma} &~\frac{\displaystyle\sum_{i=1}^M\!C_i\frac{\Lambda_i^3}{\gamma_i}\!+\!\sum_{i=1}^M \!D_i\frac{\Lambda_i^3}{\gamma_i^2}\!+\!\sum_{i=1}^M\!\sum_{j\neq i}\!G_{i,j}\frac{\Lambda_i^2\Lambda_j^2}{\gamma_i\gamma_j}}{f(\boldsymbol\Lambda,\boldsymbol\gamma)}\label{opt:Bernoulli_AF_relax_a}\\
\mbox{subject to}&~\sum_{i=1}^M\Lambda_i\leq\bar{\Lambda},\\
&~0<\Lambda_i{\black\leq\epsilon_{\Lambda}}~\text{and } \gamma_i\geq \delta~, \text{for } i=1,\ldots,M.
\end{align}
\end{subequations}
Notice that the relaxed problem is still non-convex, but can be approximated by a series of geometric programming (GP) problems using the condensation method  \cite{Boyd2007}.

\begin{algorithm}[t]
\caption{Optimized Sensor {\black Densities} and Optimized Energy Thresholds ({\black OI-OET}) Scheme in the AF System}\label{alg:AF_CM}
    \begin{enumerate}
        \item[] {\bf Initialization:} Set $\ell=0$ and a solution accuracy $\epsilon>0$. {Find a feasible solution $\left(\tilde{\boldsymbol\Lambda}^{(0)},\tilde{\boldsymbol\gamma}^{(0)}\right)$.}
        \item[] {\bf Iteration:}\\[-0.4cm]
        \begin{enumerate}
        \item[(i)] Compute $\beta_{i,j}^{(\ell)}$, for $i,j=1,\ldots,M$, by \eqref{eq:beta_ij}.
        \item[(ii)] Replace $f(\boldsymbol\Lambda,\boldsymbol\gamma)$ with $\hat f^{(\ell)}(\boldsymbol\Lambda,\boldsymbol\gamma)$ and solve \eqref{opt:Bernoulli_AF_relax}. Let the solution be            $\left(\tilde{\boldsymbol\Lambda}^{(\ell+1)},\tilde{\boldsymbol\gamma}^{(\ell+1)}\right)$.
  \item[(iii)] Repeat (i) and (ii) until $$\frac{|J_{\rm  AF}(\tilde{\boldsymbol\Lambda}^{(\ell+1)},\tilde{\boldsymbol\gamma}^{(\ell+1)})-J_{\rm AF}(\tilde{\boldsymbol\Lambda}^{(\ell)},\tilde{\boldsymbol\gamma}^{(\ell)})|}{J_{\rm AF}(\tilde{\boldsymbol\Lambda}^{(\ell)},\tilde{\boldsymbol\gamma}^{(\ell)})}\leq\epsilon.$$
    \end{enumerate}
    \end{enumerate}
\end{algorithm}

Specifically, in each iteration of the condensation method, the function $f(\boldsymbol\Lambda,\boldsymbol\gamma)$ is replaced by its monomial approximation so that {\black the objective function in \eqref{opt:Bernoulli_AF_relax_a}} becomes a posynomial and that the problem in \eqref{opt:Bernoulli_AF_relax} can be formulated as a standard GP problem.  More specifically, let $\left(\tilde{\boldsymbol\Lambda}^{(\ell)},\tilde{\boldsymbol\gamma}^{(\ell)}\right)$ be the solution obtained in the $\ell$-th iteration of the condensation method. Then, based on the inequality between arithmetic and geometric means \cite{Chiang2007}, it follows that
\begin{align}
f(\boldsymbol\Lambda,\boldsymbol\gamma)=\sum_{i=1}^M\sum_{j=1}^M \frac{A_{i,j}\Lambda_i^2\Lambda_j^2}{\gamma_i\gamma_j}\geq \prod_{i=1}^M\prod_{j=1}^M\!\left(\!\frac{\frac{A_{i,j}\Lambda_i^2\Lambda_j^2}{\gamma_i\gamma_j}}{\beta_{i,j}^{(\ell)}}\!\right)^{\!\!\!\beta_{i,j}^{(\ell)}}\triangleq\hat f^{(\ell)}(\boldsymbol\Lambda,\boldsymbol\gamma),\label{eq:f_hat}
\end{align}
where
\begin{equation}\label{eq:beta_ij}
\beta_{i,j}^{(\ell)}=\left.A_{i,j}\frac{(\tilde\Lambda_i^{(\ell)})^2(\tilde\Lambda_j^{(\ell)})^2}{\tilde\gamma_i^{(\ell)}\tilde\gamma_j^{(\ell)}}\right/f\left(\tilde{\boldsymbol\Lambda}^{(\ell)},\tilde{\boldsymbol\gamma}^{(\ell)}\right).
\end{equation}
By approximating $f(\boldsymbol\Lambda,\boldsymbol\gamma)$ with $\hat f^{(\ell)}(\boldsymbol\Lambda,\boldsymbol\gamma)$, the optimization problem in \eqref{opt:Bernoulli_AF_relax} becomes a standard GP, {\black which can be converted to a convex optimization problem and solved using the interior point method \cite{Boyd2004}.}
%The complexity required to solve such a problem depends primarily on the number of constraints and, thus, scales as $O((2M+1)^{3.5})$ for the problem in \eqref{opt:Bernoulli_AF_relax} \cite{Boyd2004,Rashtchi2014}.}
Note that solving the problem given in \eqref{opt:Bernoulli_AF_relax} by replacing $f(\boldsymbol\Lambda,\boldsymbol\gamma)$ with $\hat f^{(\ell)}(\boldsymbol\Lambda,\boldsymbol\gamma)$ yields a solution that is also feasible in \eqref{opt:Bernoulli_AF_relax} and can be used to find the solution in the next iteration of the condensation method. By initiating with a feasible solution $(\tilde{\boldsymbol\Lambda}^{(0)},\tilde{\boldsymbol\gamma}^{(0)})$, the above process can be repeated until the objective value converges. The procedure is summarized in Algorithm \ref{alg:AF_CM}. An approximated solution to the original problem in \eqref{opt:Bernoulli_AF} is thus obtained and the optimized energy threshold is further rounded to the nearest multiple of $\delta$. The effectiveness of this scheme is demonstrated through Monte Carlo simulations in Section \ref{sec:sim}.

{\black Following \cite[Chapter 11]{Boyd2004}, the GP problem in \eqref{opt:Bernoulli_AF_relax} can be converted to a convex optimization problem with $n\triangleq M^2+4M$ parameters and $m\triangleq 2M^2+6M+1$ inequality constraints. The objective and the constraints of this problem can be further used to synthesize a log-barrier function that satisfies the self-concordant property, and Newton's method can then be used to solve it. The complexity of each Newton step grows as $O(mn^2+n^3)$ \cite{Nesterov1994} and the number of Newton steps required can be bounded by $\sqrt m$ \cite{Boyd2004}. Hence, the computational complexity of the problem grows as $O(M^7)$. Details are omitted due to space limitations, but can be obtained following the steps in \cite{Boyd2004}. }

\subsection{Complexity Reduction}

%{\black The complexity of solving geometric programming such as problem \eqref{opt:Bernoulli_AF_relax} is $O((2M+1)^{3.5})$, where $2M+1$ is the number of inequality constraints\cite{Rashtchi2014}.
It is necessary to note that, even though the above approach can yield good solutions to the random sensor deployment problem, the complexity can be high when the number of subregions, $M$, is large. This is not a problem in most cases since the sensor densities and energy thresholds need only be computed offline. However, if it is necessary to reduce the complexity without reducing the resolution of the subregions, one can further reduce the number of parameters by assuming that the parameters in neighboring subregions are the same. This assumption is reasonable since the dimensions of a subregion is assumed to be much smaller than the spatial variations of the sensor field and the energy arrivals.

Suppose that the $M$ subregions are combined into $N$ clusters $\calC_1,\calC_2,\ldots,\calC_N$, each consisting of $M_C=M/N$ subregions.
%Thus, we have $M=M_CN$.
Let $\boldsymbol\Lambda_C=[\Lambda_{(1)},\Lambda_{(2)},\ldots,\Lambda_{(N)}]$ and $\boldsymbol\gamma_C=[\gamma_{(1)},\gamma_{(2)},\ldots,\gamma_{(N)}]$, where $\Lambda_{(n)}$ and $\gamma_{(n)}$ are the values of $\Lambda_i$ and $\gamma_i$, respectively, for all $i\in\calC_n$. Then, by letting $A_{(n),(m)}\triangleq\sum_{i\in\calC_n}\sum_{j\in\calC_m}A_{i,j}$, $C_{(n)}\triangleq\sum_{i\in\calC_n}C_i$, $D_{(n)}\triangleq\sum_{i\in\calC_n}D_i$, and $G_{(n),(m)}\triangleq\sum_{i\in\calC_n}\sum_{j\in\calC_m,i\neq j}G_{i,j}$, the optimization problem in \eqref{opt:Bernoulli_AF_relax} can be reduced %to the following:
\begin{subequations}\label{opt:reduced_complexity_AF}
\begin{align}
\min_{\boldsymbol\Lambda_C,\boldsymbol\gamma_C} \hspace{.5cm} & \hspace{-.5cm}~\frac{\displaystyle\sum_{n=1}^N\!\!\left[\!C_{(n)}\frac{\Lambda_{(n)}^3}{\gamma_{(n)}}\!+\!\!D_{(n)}\frac{\Lambda_{(n)}^3}{\gamma_{(n)}^2}\!+\!\!\sum_{m=1}^N \!\!G_{(n),(m)}\frac{\Lambda_{(n)}^2\Lambda_{(m)}^2}{\gamma_{(n)}\gamma_{(m)}}\right]}{f_C(\boldsymbol\Lambda_C,\boldsymbol\gamma_C)}\\
\mbox{subject to}&~\sum_{n=1}^NM_C\Lambda_{(n)}\leq\bar{\Lambda},\\
&~0<\Lambda_{(n)}{\black\leq\epsilon_{\Lambda}},~\gamma_{(n)}\geq \delta~, ~n=1,\ldots,N.
\end{align}
\end{subequations}
The problem can then be solved using the condensation method, similar to that in Algorithm \ref{alg:AF_CM}{\black , and the complexity is reduced to $O((M/M_C)^{7})$.}

%%%%%%%%%%%%%%%%%%%%%%%%%%%%%%%%%%%%%%%%%%%%%%%%%%%%%%%%%%%%%%%%%%%%%%%%%%%%%%%%%%%%%%%%%%%%%%%%%%%%%%%%%%%%%%%%%%%%
%% Optimized Sensor Deployment, Transmission Control, and Bit Allocation Strategies for Digital Forwarding Systems %
%%%%%%%%%%%%%%%%%%%%%%%%%%%%%%%%%%%%%%%%%%%%%%%%%%%%%%%%%%%%%%%%%%%%%%%%%%%%%%%%%%%%%%%%%%%%%%%%%%%%%%%%%%%%%%%%%%%%
\section{Optimized Sensor {\black Densities} and Energy Thresholds
%Deployment and Transmission Control
for Digital Forwarding Systems}\label{sec:Opt_Deploy_and_TC_DF}
\begin{figure}[t]
     \centering
     \includegraphics[scale=0.6]{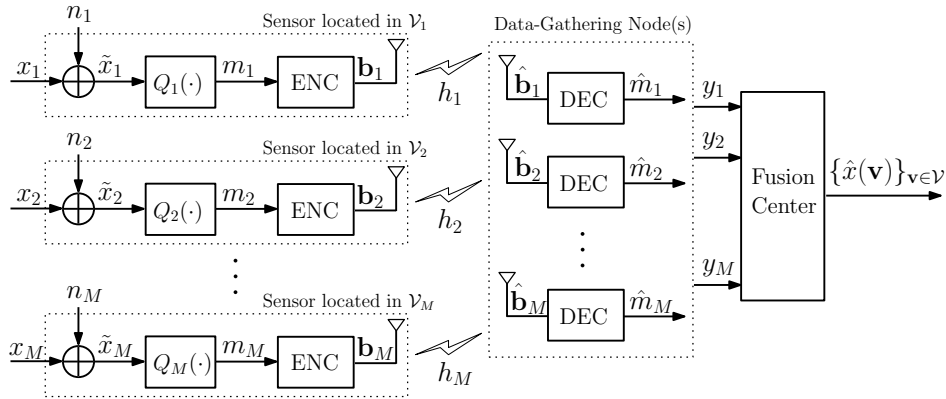}
     \caption{Illustration of filed reconstruction for the digital forwarding system.}
     \label{fig:system_model_DF}
\end{figure}
In this section, we optimize the sensor {\black densities} $\{\lambda_i\}_{i=1}^M$ and the energy thresholds $\{\gamma_i\}_{i=1}^M$ in DF systems based on the minimization of the average MSE upper bound in \eqref{eq:avMSEupper_general}.

In the DF system, each sensor first quantizes its local measurement into a binary representation vector and forwards it to the destination, where the measurement is reconstructed. An illustration of the DF system is given in Fig. \ref{fig:system_model_DF}. Suppose that $B_i$ is the number of quantization bits at a sensor in subregion $\calV_i$. The sensor measurements at the sensor are quantized into $2^{B_i}$ representation levels given by the set $\mathcal{M}_i\triangleq\{m_{i,1},\ldots, m_{i,2^{B_i}}\}$ using the quantization function $Q_i$ defined by $Q_i(\tilde x_i)=\min\{m_i\in\mathcal{M}_i: |\tilde x_i-m_i|\leq |\tilde x_i-m_i'|, \forall m_i'\in\mathcal{M}_i\}$. The index of the representation level is then encoded into the binary vector ${\bf b}_i$ and transmitted to the closest data-gathering node. Suppose that $m_i=Q_i(\tilde x_i)$ is the quantized value of $\tilde x_i$ at the sensor in $\calV_i$ and $\hat m_i$ is the corresponding value reconstructed at the  data-gathering node based on its received signal. In this case,  the effective received signal at the data-gathering node can be written as $y_i=\hat m_io_i$ and, together with the quantized measurements from other sensors, is utilized to perform the LMMSE estimate at the fusion center. The MSE and its upper bound can be written similarly as \eqref{eq:avMSE} and \eqref{eq:avMSEupper_general}, respectively. Notice that the MSE is affected by both the quantization error that appears when representing $\tilde x_i$ with $m_i$ and the channel error that causes the difference between $m_i$ and $\hat m_i$.

Following the procedure in \cite{Chaudhary2012}, we adopt a uniform quantizer, where $m_{i,l}=\frac{(2l-1-2^{B_i})\Delta_{Q_i}}{2}$, for $l=1\ldots,2^{B_i}$, and $\Delta_{Q_i}=\frac{2W}{2^{B_i}-1}$, for $W$ chosen sufficiently large such that $\Pr(|\tilde{x}_i|\geq W)\approx0$. In this case, the quantization error $\epsilon_i=m_i-\tilde x_i$ is bounded in $[-\frac{\Delta_{Q_i}}{2},\frac{\Delta_{Q_i}}{2}]$ and can be approximated as a uniform random variable over this region when $\Delta_{Q_i}$ is sufficiently small. In this case, the variance of $\epsilon_i$ is given by $\sigma_{\epsilon_i}^2=\frac{W^2}{3(2^{B_i}-1)^2}$. Moreover, it has been shown in  \cite{Chaudhary2012} and \cite{Widrow2008} that, for $B_i$ sufficiently large, for all $i$, the quantization errors $\{\epsilon_i\}_{i=1}^M$ are approximately uncorrelated and independent of the sensor measurements $\{\tilde x_i\}_{i=1}^M$. Once the quantized value $m_i$ is obtained, it is converted into a $B_i$-bit vector ${\bf b}_i$ using a natural binary code (where ${\bf b}_i$ is taken as the binary representation of the quantization index) and transmitted to the data-gathering node. By adopting BPSK modulation \cite{Chaudhary2012}, the bit-error probability of the transmission is $\varepsilon_{i}\!=\!\calQ(\sqrt{\frac{h_i^2e_i}{\sigma_w^2B_i}})$, where $h_i$ is the channel coefficient to the closest data-gathering node, $e_i/B_i$ is the energy per bit, and $\calQ(u)\!=\!\frac{1}{\sqrt{2\pi}}\int_u^{\infty}e^{-\frac{u^2}{2}}du$. By assuming that $\hat{\bf b}_i=[\hat b_{i,1},\ldots, \hat b_{i,B_i}]$ is the binary vector received by the data-gathering node, the reconstructed quantization level is then $\hat m_i=[2(\sum_{k=1}^{B_i}{ \hat b_{i,k}}2^{B_i-k}+1)-1-2^{B_i}]\frac{\Delta_{Q_i}}{2}$ since $\sum_{k=1}^{B_i}{ \hat b_{i,k}}2^{B_i-k}+1$ is the integer value of the binary vector $\hat{\bf b}_i$. Properties of the quantization and channel errors are utilized to obtain explicit expressions of the MSE upper bound.

In obtaining a tractable upper bound, let $\bar\by\triangleq[\bar y_1,\ldots, \bar y_M]^T$, where $\bar y_i=m_io_i$, be the vector of received signals at the data-gathering node when the channel is noiseless. By also letting $\tilde\by=[\tilde y_1,\ldots,\tilde y_M]^T$, where $\tilde y_i\triangleq\bar y_i-y_i$, the MSE upper bound in \eqref{eq:avMSEupper_min} can be further upper-bounded as
\begin{align}
\bar\xi_{\rm upper, DF}&= \frac{1}{|\mathcal V|}\int_{\mathbf v\in\mathcal V}\min_{\bk\in\mathbb{R}^M}E[|x(\bv)-\bk^T\bar\by+\bk^T\bar\by-\bk^T\by|^2]d\bv\\
&{\black\leq\frac{2}{|\mathcal V|}\int_{\mathbf v\in\mathcal V}\min_{\bk\in\mathbb{R}^M}\left\{E\left[| x(\bv)- \bk^T\bar\by|^2\right]+E\left[|\bk^T\bar\by-\bk^T\by|^2\right]\right\}d\bv\label{eq:avMSEupper_DF1}}\\
&{\black\leq2\left[\sigma_x^2-\frac{\left(\tr\left(\bPhi\bD_\balpha\bD_\balpha\right)\right)^2}{\tr\left(\bPhi\bD_\balpha(\bC_{\bar\by\bar\by}+\bC_{\tilde\by\tilde\by})\bD_\balpha\right)}\right]\label{eq:avMSEupper_DF2}}\\
&\triangleq\bar\xi_{\rm obj,DF}(\blambda,\bgamma), \label{eq:_xi_obj}
\end{align}
where $\bC_{\bar\by\bar\by}\triangleq E[\bar\by\bar\by^T]$ and $\bC_{\tilde\by\tilde\by}\triangleq E\left[\tilde\by\tilde\by^T\right]$. {\black The first inequality follows from the fact that $(a+b)^2\leq 2(a^2+b^2)$ and the second inequality is shown in Appendix \ref{APP:xi_obj_DF}. The first inequality splits the MSE into two terms, one contributed by quantization error and the other by channel noise. The bound is tight when $a$ is approximately equal to $b$, that is, when the two MSE contributions are approximately the same. Furthermore, the upper bound in \eqref{eq:avMSEupper_DF2} follows similar to \eqref{eq:avMSEupper_min_given_o} and \eqref{eq:avMSEupper2pre} and, thus, are tight under similar conditions. Similar to the AF case, even though the upper bounds may not be tight in general, they exhibit similar behaviors as the actual MSE with respect to the sensor densities and transmission thresholds and, thus, are used as objective functions in our problem.}
%{\black(Please see the derivation in APPENDIX \ref{APP:xi_obj_DF}) as
%\begin{align}
%\bar\xi_{\rm upper, DF}&\leq2\left[\sigma_x^2-\frac{\left(\tr\left(\bPhi\bD_\balpha\bD_\balpha\right)\right)^2}{\tr\left(\bPhi\bD_\balpha(\bC_{\bar\by\bar\by}+\bC_{\tilde\by\tilde\by})\bD_\balpha\right)}\right]\\
%&\triangleq\bar\xi_{\rm obj,DF}(\blambda,\bgamma), \label{eq:_xi_obj}
%\end{align}
%where $\tilde\by=[\tilde y_1,\ldots,\tilde y_M]^T$ with $\tilde y_i\triangleq\bar y_i-y_i$, $\bC_{\bar\by\bar\by}\triangleq E[\bar\by\bar\by^T]$ and $\bC_{\tilde\by\tilde\by}\triangleq E\left[\tilde\by\tilde\by^T\right]$.

To further analyze $\bar\xi_{\rm obj,DF}(\blambda,\bgamma)$ in \eqref{eq:avMSEupper_DF2}, we note that the elements in $\bC_{\bar \by\bar\by}$ and $\bC_{\tilde\by\tilde\by}$ can be derived as
\begin{equation}
\{\bC_{\bar\by\bar\by}\}_{i,j}\!=\!\left\{\!\!\!\begin{array}{cc}(\sigma_x^2+\sigma_n^2+\sigma_{\epsilon_i}^2)\alpha_i, & \text{for } i=j\\
\sigma_x^2\rho(\bv_i,\bv_j)\alpha_i\alpha_j, & \text{for } i\neq j,\end{array}\right.
\end{equation}
and
\begin{align}
\{\!\bC_{\tilde\by\tilde\by}\!\}_{i,j}\!=\!\!\left\{
\begin{array}{ll}\!\!\!\!2^{2B_i}\!\Delta_{Q_i}^2q_i\!\left(\frac{1-4^{-B_i}}{3}\!+\!q_i\eta_{i,i}\right)\!\alpha_i, &\!\!\! \text{for } i\!=\!j\\
\!\!\!\!2^{B_i+B_j}\!\Delta_{Q_i}\!\Delta_{Q_j}\!q_iq_j\eta_{i,j}\alpha_i\alpha_j, &\!\!\! \text{for } i\!\neq\! j,\end{array}\right.
\end{align}
where
\begin{align}
q_i\triangleq\Pr(\hat b_{i,k}=0|b_{i,k}=1,o_i=1),
\end{align}
\begin{align}
\eta_{i,j}\triangleq\left\{
\begin{array}{cc}\sum_{k=1}^{B_i}\sum_{l\neq k}\pi_{\{i,k\},\{i,l\}}2^{-k}2^{-l}, & \text{for } i=j\\
\sum_{k=1}^{B_i}\sum_{l=1}^{B_j}\pi_{\{i,k\},\{j,l\}}2^{-k}2^{-l}, & \text{for } i\neq j,\end{array}\right.
\end{align}
with
\begin{align}
\pi_{\{i,k\},\{j,l\}}&\!\triangleq\! 1\!-\!2[\Pr(b_{i,k}\!=\!1,b_{j,l}\!=\!0|o_i\!=\!1,o_j\!=\!1)\!+\!\Pr(b_{i,k}\!=\!0,b_{j,l}\!=\!1|o_i\!=\!1,o_j\!=\!1)]\label{eq:pi}
\end{align}
for $k=1,\ldots,B_i$, $l=1,\ldots,B_j$, $i,j=1,\ldots,M$. The derivation of $\bC_{\tilde\by\tilde\by}$ and a more explicit expression of  $\pi_{\{i,k\},\{j,l\}}$ can be found in Appendices \ref{APP:C_tilde_y} and \ref{APP:pi}, respectively.
By taking $\bar\xi_{\rm obj,DF}(\blambda,\bgamma)$ as the objective function, the search for the optimal sensor {\black densities} and energy thresholds can be formulated as
\begin{subequations}\label{eq:problem_formulation_DF}
\begin{align}
\min_{\boldsymbol\lambda,\boldsymbol\gamma} \quad&\bar\xi_{\rm obj, DF}(\boldsymbol\lambda,\boldsymbol\gamma)\\
\mbox{subject to}\quad&\sum_{i=1}^M\lambda_i\Delta\leq\bar\Lambda,\\
&0<\lambda_i\Delta{\black\leq\epsilon_{\Lambda}},~\text{for }i=1,\ldots, M.
\end{align}
\end{subequations}

For the reasons that will be evident later, we perform the change of variables where $\lambda_i$ is replaced with $\alpha_i=\bar{F}_{e_i}(\gamma_i)\lambda_i\Delta$, for all $i=1,\ldots, M$. Then, by omitting the terms that are not relevant to the optimization, the problem can be written as
\begin{subequations}
\begin{align}
\min_{\balpha,\bgamma} &~J_{\rm DF}(\balpha,\bgamma)\\
\mbox{subject to}&~\sum_{i=1}^M\frac{\alpha_i}{\bar{F}_{e_i}(\gamma_i)}\leq\bar\Lambda,\\
&~0<\alpha_i\leq {\black\epsilon_\Lambda}\bar{F}_{e_i}(\gamma_i), ~\text{for }i=1,\ldots, M,
\end{align}
\end{subequations}
where
\begin{align}
J_{\rm DF}(\balpha,\bgamma)&=\left(\frac{\left(\tr\left(\bPhi\bD_\balpha\bD_\balpha\right)\right)^2}{\tr\left(\bPhi\bD_\balpha(\bC_{\bar\by\bar\by}+\bC_{\tilde\by\tilde\by})\bD_\balpha\right)}\right)^{-1}.
\end{align}

Notice that, in $J_{\rm DF}(\balpha,\bgamma)$, the dependence on $\balpha$ lies in $\bD_\balpha$, $\bC_{\bar\by\bar\by}$, and $\bC_{\tilde\by\tilde\by}$ whereas the dependence on $\bgamma$ lies only in $\bC_{\tilde\by\tilde\by}$. The optimal solution of this problem is still difficult to find due to the non-convexity of the problem. However, an approximate solution can be found by using an alternating optimization algorithm \cite{Niesen2009}, where $\balpha$ and $\bgamma$ are optimized in turn while keeping the other fixed and the process is repeated iteratively until there is no appreciable decrease in the objective function.
%{\black Although the solution obtained from alternating optimization algorithm is not a global optimum in general, it}
{\black The algorithm is guaranteed to converge since the objective is bounded below and is minimized in each step of the algorithm, but may converge to only a local minimum in general.} Details of the optimization of $\balpha$ and $\bgamma$ are described in the following subsection using the Bernoulli energy arrival model as an example.

\subsection{Solution for the Bernoulli Energy Arrival Case}\label{sec:Bernoulli_DF}

Recall that, in the Bernoulli energy arrival case, the energy threshold $\gamma_i$ can be set, without loss of generality, as a multiple of $\delta$. In this case, we have $q_i=\calQ\left(\sqrt{\frac{h_i^2\gamma_i}{\sigma_w^2B_i}}\right)$ and $\bar{F}_{e_i}(\gamma_i)=p_i\delta/\gamma_i$. By relaxing the integer constraint on $\gamma_i$ into the linear constraint $\gamma_i\geq\delta$, the optimization problem becomes
\begin{subequations}\label{eq:problem_formulation_DF_relaxed}
\begin{align}
\min_{\balpha,\bgamma} &~J_{\rm DF}(\balpha,\bgamma)=\frac{J^{\text{num}}_{\rm DF}(\balpha,\bgamma)}{J^{\text{den}}_{\rm DF}(\balpha,\bgamma)}\\
\mbox{subject to}&~\sum_{i=1}^M\frac{\alpha_i\gamma_i}{p_i\delta}\leq\bar\Lambda,\\
&~0<\alpha_i\leq {\black\epsilon_\Lambda}p_i\delta/\gamma_i,~\text{for }i=1,\ldots, M,\\
&~\gamma_i\geq\delta,~\text{for }i=1,\ldots, M,
\end{align}
\end{subequations}
where the term in the numerator is
\begin{align}
J^{\text{num}}_{\rm DF}(\balpha,\!\bgamma)\!=\!\sum_{i=1}^M\phi_{i,i}\!\left[\sigma_x^2\!+\!\sigma_n^2\!+\!\sigma^2_{\epsilon_i}\!+\!\zeta_{i,i}(\gamma_i)\right]\alpha_i^3+\!\sum_{i=1}^M\!\sum_{j\neq i}\!\phi_{i,j}\!\!\left[\sigma_x^2\rho(\bv_i,\!\bv_j)\!+\!\zeta_{i,j}(\gamma_i,\!\gamma_j)\right]\!\alpha_i^2\alpha_j^2\label{eq.DF_Bernoulli_num}
\end{align}
with
\begin{align}
\zeta_{i,i}(\gamma_i)\!\triangleq \!\frac{4W^24^{B_i}}{(2^{B_i}\!-\!1)^2}\calQ\!\left(\!\!\sqrt{\!\frac{h_i^2\gamma_i}{\sigma_w^2B_i}}\right)\!\!\left[\!\frac{1\!-\!4^{-B_i}}{3}\!+\!\calQ\!\left(\!\!\sqrt{\!\frac{h_i^2\gamma_i}{\sigma_w^2B_i}}\right)\!\eta_{i,i}\!\right]
\end{align}
and
\begin{align}
\zeta_{i,j}(\gamma_i,\gamma_j)\triangleq\! \frac{4W^22^{B_i+B_j}}{(2^{B_i}\!-1)(2^{B_j}\!-1)}\calQ\!\left(\!\sqrt{\frac{h_i^2\gamma_i}{\sigma_w^2B_i}}\right)\!\calQ\!\left(\!\sqrt{\frac{h_j^2\gamma_j}{\sigma_w^2B_j}}\right)\!\eta_{i,j},
\label{eq.DF_Bernoulli_zeta}\end{align} for $i\neq j$,
and the term in the denominator is
\begin{align}
J^{\text{den}}_{\rm DF}(\balpha,\bgamma)=\sum_{i=1}^M\sum_{j=1}^M\phi_{i,i}\phi_{j,j}\alpha_i^2\alpha_j^2.\label{eq.DF_Bernoulli_den}
\end{align}

Let $\balpha^{(\ell)}$ and $\bgamma^{(\ell)}$ be the solutions obtained in the $\ell$-th iteration of the proposed algorithm. In the $(\ell+1)$-th iteration, we perform the optimization in the following two steps.

\underline{\bf Step 1 (Optimization of $\balpha$):} In Step 1, we first find the optimal value of $\balpha$ given $\bgamma=\bgamma^{(\ell)}$. That is, we find
\begin{equation}
\balpha^{(\ell+1)}=\mathop{\arg\min}_{\balpha\in\calF_\balpha^{(\ell+1)}} J_{\rm DF}(\balpha,\bgamma^{(\ell)}),\label{opt:alpha}
\end{equation}
where
\begin{align}
\calF_\balpha^{(\ell+1)}\!=\!\left\{\!\balpha\!\left|\sum_{i=1}^M\frac{\gamma_i^{(\ell)}}{p_i\delta}\alpha_i\leq\bar{\Lambda} \text{ and }0<\alpha_i\leq {\black\epsilon_\Lambda}\frac{p_i\delta}{\gamma_i^{(\ell)}},\;\forall i\right.\!\right\}
\end{align}
is the constraint set on $\balpha$ in iteration $\ell+1$. By \eqref{eq.DF_Bernoulli_num}-\eqref{eq.DF_Bernoulli_zeta}, the optimization problem in \eqref{opt:alpha} can be written explicitly as
\begin{subequations}\label{opt:Bernoulli_alpha}
\begin{align}
\min_{\balpha} \quad&\frac{\sum_{i=1}^M\tilde C_i^{(\ell)}\alpha_i^3+\sum_{i=1}^M\sum_{j\neq i}\tilde D_{i,j}\alpha_i^2\alpha_j^2}{\sum_{i=1}^M\sum_{j=1}^M\tilde A_{i,j}\alpha_i^2\alpha_j^2}\label{opt:Bernoulli_alpha_a}\\
\mbox{subject to}\quad&\sum_{i=1}^M\frac{\gamma_i^{(\ell)}}{p_i\delta}\alpha_i\leq\bar\Lambda,\\
&0<\alpha_i\leq {\black\epsilon_\Lambda}p_i\delta/\gamma_i^{(\ell)},~\text{for }i=1,\ldots, M,
\end{align}
\end{subequations}
where $\tilde A_{i,j}\triangleq\phi_{i,i}\phi_{j,j}$, $\tilde C_i^{(\ell)}\triangleq\phi_{i,i}[\sigma_x^2+\sigma_n^2+\sigma_{\epsilon_i}^2+\zeta_{i,i}(\gamma_i^{(\ell)})]$ and $\tilde D_{i,j}\triangleq \phi_{i,j}[\sigma_x^2\rho(\bv_i,\bv_j)+\zeta_{i,j}(\gamma_i^{(\ell)},\gamma_j^{(\ell)})]$.
The optimization problem is nonconvex but can be solved approximately using the condensation method, similar to that in Algorithm \ref{alg:AF_CM}, with monomial approximations of the denominator of \eqref{opt:Bernoulli_alpha_a}.

\underline{\bf Step 2 (Optimization of $\bgamma$):} In Step 2, we find the optimal value of $\bgamma$ given $\balpha=\balpha^{(\ell+1)}$, i.e., we find
\begin{equation}
\bgamma^{(\ell+1)}=\mathop{\arg\min}_{\bgamma\in\calF_\bgamma^{(\ell+1)}} J_{\rm DF}(\balpha^{(\ell+1)},\bgamma),
\end{equation}
where
\begin{align}
\calF_\bgamma^{(\ell+1)}\!=\!\left\{\!\bgamma\!\left|\sum_{i=1}^M\!\frac{\alpha_i^{(\ell+1)}}{p_i\delta}\gamma_i\!\leq\!\bar{\Lambda} \text{ and }\delta\!\leq\!\gamma_i\!\leq \!\!\frac{{\black\epsilon_\Lambda}p_i\delta}{\alpha_i^{(\ell+1)}},\forall i\right.\!\!\right\}\label{eq.F_gamma}
\end{align}
is the constraint set on $\bgamma$ in iteration $\ell+1$. Since $J_{\rm DF}(\balpha,\bgamma)$ depends on $\bgamma$ only through $\{\zeta_{i,j}\}_{i,j=1}^M$, the optimization over $\bgamma$ can be formulated equivalently as
\begin{align}\label{eq:opt_gamma_DF}
\min_{\bgamma\in\calF_\bgamma^{(\ell+1)}} \sum_{i=1}^M\phi_{i,i}\zeta_{i,i}(\gamma_i)\left(\!\alpha_i^{(\ell+1)}\!\right)^3\!+\!\sum_{i=1}^M\!\sum_{j\neq i}\phi_{i,j}\zeta_{i,j}(\gamma_i,\gamma_j)\left(\!\alpha_i^{(\ell+1)}\!\right)^2\left(\!\alpha_j^{(\ell+1)}\!\right)^2.
\end{align}
Notice that the objective function is not a convex function of $\bgamma$. However, by applying the upper bound $\calQ(u)\leq \frac{1}{2}\exp\left(-\frac{u^2}{2}\right)$, for $u>0$, and by taking the high SNR approximation,
we have
\begin{equation}\label{eq:zeta_upper1}
\zeta_{i,i}(\gamma_i)\lesssim\frac{2W^2(2^{B_i}+1)}{3(2^{B_i}-1)}e^{-\frac{h_i^2\gamma_i}{2\sigma_w^2B_i}}
\end{equation}
and
\begin{align}\label{eq:zeta_upper2}
\zeta_{i,j}(\gamma_i,\gamma_j)\!\lesssim\!
\frac{W^22^{B_i+B_j}}{(2^{B_i}-1)(2^{B_j}-1)}\eta_{i,j}e^{-\frac{\frac{h_i^2\gamma_i}{B_i}+\frac{h_j^2\gamma_j}{B_j}}{2\sigma_w^2}},
\end{align}
for $i\neq j$. Then, the optimization problem in \eqref{eq:opt_gamma_DF} can then be approximated as the problem below:
\begin{subequations}\label{eq:opt_gamma_upper}
\begin{align}
\notag\min_{\boldsymbol\gamma} &~\sum_{i=1}^M\frac{\phi_{i,i}\left(\alpha_i^{(\ell+1)}\right)^32(2^{B_i}+1)}{3(2^{B_i}-1)}e^{-\frac{h_i^2\gamma_i}{2\sigma_w^2B_i}}\\
&\!+\!\!\sum_{i=1}^M\!\sum_{j\neq i}\!\!\frac{\phi_{i,j}\!\!\left(\!\alpha_i^{(\ell+1)}\alpha_j^{(\ell+1)}\!\right)^{\!\!2}\!\!2^{B_i+B_j}}{(2^{B_i}-1)(2^{B_j}-1)}\eta_{i,j}e^{-\frac{\frac{h_i^2\gamma_i}{B_i}+\frac{h_j^2\gamma_j}{B_j}}{2\sigma_w^2}}\\
\mbox{subject to}&~\sum_{i=1}^M\frac{\alpha_i^{(\ell+1)}}{p_i\delta}\gamma_i\leq\bar{\Lambda}\\
&\delta\leq\gamma_i\leq {\black\epsilon_\Lambda}p_i\delta/\alpha_i^{(\ell+1)},~\text{for }i=1,\ldots,M,
\end{align}
\end{subequations}
%which can be
which is convex and can be solved efficiently using standard numerical approaches, such as the interior point method \cite{Boyd2004}.

\begin{algorithm}[t]
\caption{Alternating Optimization Algorithm of $\balpha$ and $\bgamma$}
%\centering
\begin{enumerate}
        \item[] {\bf Initialization:} Set $\ell=0$, $\epsilon>0$ and find a feasible initial solution $(\balpha^{(0)},\bgamma^{(0)})$.
        \item[] {\bf Iteration $\ell+1$:}\\[-0.4cm]
\begin{enumerate}
\item[(i)] Find
$\balpha^{(\ell+1)}$ by solving \eqref{opt:Bernoulli_alpha} in Step 1 using the condensation method.
\item[(ii)] Find
$\bgamma^{(\ell+1)}$ by solving \eqref{eq:opt_gamma_upper} in Step 2.
\item[(iii)] Repeat (i)-(ii) until $$\frac{|J_{\rm DF}(\balpha^{(\ell+1)},\bgamma^{(\ell+1)})-J_{\rm DF}(\balpha^{(\ell)},\bgamma^{(\ell)})|}{J_{\rm DF}(\balpha^{(\ell)},\bgamma^{(\ell)})}\leq \epsilon.$$  Take $\balpha^{(\ell+1)}$ and $\bgamma^{(\ell+1)}$ as the desired solution.
\end{enumerate}
\end{enumerate}
\label{alg:alternating}
\end{algorithm}

By alternating between the optimization problems in Steps 1 and 2 until convergence, the desired approximate solution of $\balpha$ and $\bgamma$ can be obtained. The alternating optimization algorithm is summarized in Algorithm \ref{alg:alternating}.

{\black
\subsection{Extension to DF Systems with Parity Check Bits}\label{sec:parity_bit}

In this section, we have so far investigated DF systems where raw bits are transmitted for field reconstruction. In practice, the addition of parity bits to the transmitted signal is often considered to allow for error detection at the data-gathering node. In this case, the message transmitted by the sensor can be treated as an erasure if an error has been detected and as error-free, otherwise. The sensor deployment strategy can then be derived similarly in this case.

Specifically, let us consider the simple case where only an even parity is used. In this case, the bit sequence transmitted by a sensor in subregion $\calV_i$ can be written as ${\check{\bf b}}_i=[\mathbf b_i~b_{i,{B_i+1}}]$, where $b_{i,{B_i+1}}$ is chosen such that $\sum_{k=1}^{B_i+1} b_{i,k}$ is even and is referred to as the even parity bit. By assuming that an error is always detected and treated as an erasure when it occurs, the effective received signal at the fusion center can be written as
%In this scheme, the signals utilized for the field reconstruction could include the erroneously decoded signal. Alternatively, we can also consider a parity-check bit is added into the encoded message for transmitting to the data-gathering node. As a result, incorporating erroneously decoded signal for the field reconstruction in the DF system could be avoided. Specifically, this paper considers the even parity bit system where the transmitted signal is ${\check{\bf b}}_i=[\mathbf b_i~b_{i,{B_i+1}}]$ such that mod$\left(\sum_{k=1}^{B_i+1},2\right)=0$, where $b_{i,B_i+1}$ is the even parity bit. By assuming the probability that the number of error bits is more than one is sufficiently small, which can be neglected, we focus on the case in which there is at most one bit in error. In this case, the decoded signal can be written as
%\begin{align}
%\hat m_i=\left\{
%\begin{array}{ll}
%m_i,&\text{if mod} \left(\sum_{k=1}^{B_i+1},2\right)=0\\
%\text{erasure},&\text{if mod} \left(\sum_{k=1}^{B_i+1},2\right)=1\end{array}\right..
%\end{align}
%By dropping the decoded signal when the parity check is failed, the received signal can be represented as
\begin{align}
y_i=m_i\tilde o_i
\end{align}
where
\begin{align}
\tilde o_i=\left\{
\begin{array}{ll}
0,&\text{an error is detected}\\
o_i,&\text{otherwise.}\end{array}\right.
\end{align}

Following similar procedures as in the previous scheme, an MSE upper bound of DF systems with a one-bit parity can be derived as
\begin{align}
\bar\xi&\leq\sigma_x^2-\tr\left(\bPhi\bD_{\balpha}\bD_{\tilde\bq}\bC_{\by\by}^{-1}\bD_{\tilde\bq}\bD_{\balpha}\right)\\
&\leq\sigma_x^2-\frac{\left(\tr\left(\bPhi\bD_{\balpha}^2\bD_{\tilde\bq}^2\right)\right)^2}{\tr\left(\bPhi\bD_{\balpha}\bD_{\tilde\bq}\bC_{\by\by}\bD_{\tilde\bq}\bD_{\balpha}\right)}\triangleq\bar\xi_{\rm obj,PB}(\bLambda,\bgamma)\label{eq:xi_obj_PB}
\end{align}
where $\bD_{\tilde\bq}\triangleq\diag((1-q_1)^{B_1+1},\ldots,(1-q_M)^{B_M+1})$ and
\begin{align}
\{\bC_{\by\by}\}_{i,j}=\left\{\!\!\!\begin{array}{cc}(1\!-\!q_i)^{B_i+1}(\sigma_x^2+\sigma_n^2+\sigma_{\epsilon_i}^2)\alpha_i, & \text{for } i\!=\!j,\\
\sigma_x^2\rho(\bv_i,\!\bv_j)(1\!-\!q_i)^{B_i+1\!}(1\!-\!q_j)^{B_j+1}\alpha_i\alpha_j, & \text{for } i\!\neq\! j.\end{array}\right.
\end{align}
For sufficiently small bit error probability, i.e., for $q_i\approx0~\forall i$, the objective function can be approximated as
\begin{align}
\bar\xi_{\rm obj,PB}\approx\sigma_x^2-\frac{J^{\text{num}}_{\rm PB}(\bLambda,\bgamma)}{J^{\text{den}}_{\rm PB}(\bLambda,\bgamma)}
\end{align}
where
%\begin{align}
%J^{\text{num}}_{\rm PB}(\bLambda,\bgamma)=&\left(\sum_{i=1}^M\phi_{i,i}\alpha_i^2\right)^2\label{eq:J_PB_Num}
%\end{align}
%and
%\begin{align}
%J^{\text{den}}_{\rm PB}(\bLambda,\bgamma)&=\sum\limits_{i=1}^M\phi_{i,i}(1-q_i)^{3(B_i+1)}(\sigma_x^2+\sigma_n^2+\sigma_{\epsilon_i}^2)\alpha_i^3\\
%&~+\sum_{i=1}^M\sum_{j\neq i}\phi_{i,j}\sigma_x^2\rho(\bv_i,\bv_j)(1-q_i)^{2(B_i+1)}(1-q_j)^{2(B_j+1)}\alpha_i^2\alpha_j^2\label{eq:J_PB_Den}
%\end{align}
\begin{align}
J^{\text{num}}_{\rm PB}(\bLambda,\bgamma)=&\left(\sum_{i=1}^M\phi_{i,i}\alpha_i^2\right)^2\label{eq:J_PB_Num}
\end{align}
and
\begin{align}
J^{\text{den}}_{\rm PB}(\bLambda,\bgamma)=\sum\limits_{i=1}^M\phi_{i,i}(\sigma_x^2+\sigma_n^2+\sigma_{\epsilon_i}^2)\alpha_i^3+\sum_{i=1}^M\sum_{j\neq i}\phi_{i,j}\sigma_x^2\rho(\bv_i,\bv_j)\alpha_i^2\alpha_j^2.\label{eq:J_PB_Den}
\end{align}
By further adopting the Bernoulli energy arrival model, the optimization problem can be written explicitly as
%As the way of formulating the sensor deployment problem by using the objective function from the MSE in Section \ref{sec:Opt_Deploy_and_TC_AF} and \ref{sec:Opt_Deploy_and_TC_DF}, the sensor deployment problem for the DF system with a parity bit in Bernoulli energy arrival case can be similarly formulated by the following optimization problem
\begin{subequations}\label{opt:Bernoulli_PB}
\begin{align}
\max_{\boldsymbol\Lambda,\boldsymbol\gamma}~ &\frac{\displaystyle\sum_{i=1}^M\sum_{j=1}^M \check{A}_{i,j}\frac{\Lambda_i^2\Lambda_j^2}{\gamma_i^2\gamma_j^2}}{\displaystyle\sum_{i=1}^M\check{C}_i\frac{\Lambda_i^3}{\gamma_i^3}+\sum_{i=1}^M\sum_{j\neq i}\check{G}_{i,j}\frac{\Lambda_i^2\Lambda_j^2}{\gamma_i^2\gamma_j^2}}\\
\mbox{subject to}&~\sum_{i=1}^M\Lambda_i\leq\bar\Lambda,\\
&~0<\Lambda_i\leq\epsilon_{\Lambda}<1,\;i=1,\ldots,M,\\
&~\gamma_i=Z_i\delta,\; Z_i\in\mathbb{N},\;i=1,\ldots,M,
\end{align}
\end{subequations}
where $\check{A}_{i,j}\triangleq\phi_{i,i}\phi_{j,j}p_i^2p_j^2\delta^4$, $\check{C}_i\triangleq\phi_{i,i}(\sigma_x^2+\sigma_n^2+\sigma_{\epsilon_i}^2)p_i^3 \delta^3$ and $\check{G}_{i,j}\triangleq \phi_{i,j}\sigma_x^2\rho(\bv_i,\bv_j)p_i^2p_j^2\delta^4$. Notice that the problem is similar to that obtained in the AF case and, thus, can be solved following similar procedures as in
%}The formulated problem \eqref{opt:Bernoulli_PB} can be solved by the similar procedure of solving \eqref{opt:Bernoulli_AF} in
Section \ref{subsec:Solution_for_Bern_energy_arrival}.
}

\section{Simulations and Performance Comparisons}\label{sec:sim}

%%%%%%%%%%%%%%%%%
%% Appendix    %%
%%%%%%%%%%%%%%%%%
\appendices
{\black
\section{Equivalence of the LMMSE Estimator in \eqref{eq:LMMSE_estimator}}\label{APP:equiv}

Let $\calO_1=\{i| o_i=1\}$ and $\calO_0=\{i| o_i=0\}$ be the index sets of the subregions with and without a transmitting sensor, respectively, and let $|\calO_1|=r$ and $|\calO_0|=M-r$. Moreover, let $\bz=[z_1,\ldots, z_r]^T$ be the $r\times 1$ vector obtained by removing the entries in $\by$ that correspond to the indices in $\calO_0$. In this case, $\bC_{x(\bv)\bz}^\bo\triangleq E[x(\bv)\bz^T|\bo]$ is equivalent to the vector $\bC_{\bx(\bv)\by}^\bo$ with the entries in $\calO_0$ removed, and $\bC_{\bz\bz}^\bo\triangleq E[\bz\bz^T|\bo]$ is equivalent to the matrix $\bC_{\by\by}^\bo$ with the rows and columns in $\calO_0$ removed. The eigenvalue decomposition of $\bC_{\bz\bz}^\bo$ can be written as
\begin{align}
\bC_{\bz\bz}^\bo=\sum_{k=1}^r \varrho_k\bu_k\bu_k^T,
\end{align}
where $\varrho_k$,  for $k=1,\ldots,r$, is the $k$-th eigenvalue, labelled such that $\varrho_1\geq\varrho_2\geq\ldots\geq\varrho_r>0$, and $\bu_k=[u_{k,1},\ldots, u_{k,r}]^T$ is the corresponding eigenvector. Then, by constructing the vectors $\bv_k=[v_{k,1},\ldots, v_{k,M}]^T$, for $k=1,\ldots, r$, such that
\begin{align}
v_{k,i}=\left\{\begin{array}{ll}
u_{k,\sum_{j=1}^io_j},&\text{ if } o_i=1\\
0,&\text{ if } o_i=0,
\end{array}\right.
\end{align}
(i.e., by padding zeros into the vector $\bu_k$ at locations corresponding to the indices in $\calO_0$), we have
\begin{align}
\bC_{\by\by}^\bo=\sum_{k=1}^r \varrho_k\bv_k\bv_k^T.
\end{align}
Notice that $\bv_1, \ldots, \bv_r$ are linearly independent and, thus, are eigenvectors corresponding to the non-zero eigenvalues $\varrho_1,\ldots, \varrho_r$ of $\bC_{\by\by}^\bo$.

Therefore, the LMMSE estimator obtained by using only the received signals $r_i$, for $i\in\calO_1$, can be written as
\begin{align}
\hat x(\bv)&=\bC_{x(\bv)\bz}^\bo{\bC_{\bz\bz}^\bo}^{\!\!-1}\bz\\
&=\sum_{i=1}^r\sum_{j=1}^r\{\bC_{x(\bv)\bz}^\bo\}_i\{{\bC_{\bz\bz}^\bo}^{\!\!-1}\}_{i,j}z_j\\
&=\sum_{i=1}^r\sum_{j=1}^r\{\bC_{x(\bv)\bz}^\bo\}_i\left(\sum_{k=1}^r \varrho_k^{-1}u_{k,i} u_{k,j}\right)\!z_j\\
&=\sum_{i=1}^M\sum_{j=1}^M\{\bC_{x(\bv)\by}^{\bo}\}_i\left(\sum_{k=1}^r \varrho_k^{-1}v_{k,i}v_{k,j}\!\right)\!y_j\label{eq:u=uo}\\
&=\sum_{i=1}^M\sum_{j=1}^M\{\bC_{x(\bv)\by}^{\bo}\}_i\left\{\sum_{k=1}^r \varrho_k^{-1}\bv_k\bv_k^T\right\}_{i,j}\{\by\}_i\\
&=\bC_{x(\bv)\by}^\bo{\bC_{\by\by}^\bo}^{\!\!\dag}\by,
\end{align}
where the equality in \eqref{eq:u=uo} follows from the fact that $v_{k,i}=0,~\forall k$ for $i\in\{i|o_i=0\}$.

\section{Derivation of the Inequality in \eqref{eq:avMSEupper_DF2}}\label{APP:xi_obj_DF}

%The upper bound in \eqref{eq:avMSEupper_min} can be further upper-bounded as
%\begin{align}
%\notag\bar\xi_{\rm upper, DF}&= \frac{1}{|\mathcal V|}\int_{\mathbf v\in\mathcal V}\min_{\bk\in\mathbb{R}^M}E[\|x(\bv)-\bk^T\bar\by\\
%&\quad\qquad\qquad\qquad+\bk^T\bar\by-\bk^T\by\|^2]d\bv\\
%&\notag\leq\frac{2}{|\mathcal V|}\int_{\mathbf v\in\mathcal V}\min_{\bk\in\mathbb{R}^M}\left\{E\left[\| x(\bv)- \bk^T\bar\by\|^2\right]\right.\\
%&~\qquad\qquad\left.+E\left[\|\bk^T\bar\by-\bk^T\by\|^2\right]\right\}d\bv\label{eq:avMSEupper_DF1}
%\end{align}
%where the equality follows from the fact that $(a+b)^2\leq 2(a^2+b^2)$. L

To derive the inequality in \eqref{eq:avMSEupper_DF2}, let us define the term inside the integral of \eqref{eq:avMSEupper_DF1} as
\begin{align}
\calL(\bk,\bv)
&\triangleq \!E\!\left[|x(\bv)-\bk^T\bar\by|^2\right]\!+\!E\!\left[|\bk^T\bar\by-\bk^T\by|^2\right]\\
&=\sigma_x^2-2\bC_{x(\bv)\bar\by}\bk+\bk^T\left(\bC_{\bar\by\bar\by}+\bC_{\tilde\by\tilde\by}\right)\bk
\end{align}
where $\bC_{x(\bv)\bar\by}\triangleq E[x(\bv)\bar\by^T]$. To minimize $\calL(\bk,\bv)$, we set $\frac{\partial \calL(\bk,\bv)}{\partial\bk}=0$, which yields
\begin{equation}\label{eq:bark}
{\bf k}^\star(\bv)=\mathop{\arg\min}_{{\bf k}\in\mathbb{R}^M}\calL(\bk,\bv)=(\bC_{\bar\by\bar\by}+\bC_{\tilde\by\tilde\by})^{-1}{\bC_{x(\bv)\bar\by}}^T,
\end{equation}
and the corresponding minimum value is
\begin{align}
\calL(\bk^*(\bv),\bv)=\sigma_x^2-\bC_{x(\bv)\bar\by}(\bC_{\bar\by\bar\by}+\bC_{\tilde\by\tilde\by})^{-1}{\bC_{x(\bv)\bar\by}}^T\label{eq:Lv}
\end{align}
where
\begin{equation}\label{eq:apprCxy}
\mathbf C_{x(\bv)\bar\by}= \bC_{x(\bv)\bm}\bD_\alpha\approx\bC_{x(\bv)\bx}\bD_\alpha
\end{equation}
and $\bC_{x(\bv)\bm}\triangleq E[x(\bv)\bm^T]$ with $\bm=[m_1,\ldots, m_M]^T$ being the vector of quantized sensor measurements. The approximation in \eqref{eq:apprCxy} is made by assuming that the quantization error $\epsilon_i$ is uniformly distributed and is uncorrelated with its input $\tilde{x_i}$. By substituting \eqref{eq:Lv} into \eqref{eq:avMSEupper_DF1}, the upper bound of the average MSE becomes
\begin{align}
\bar\xi_{\rm upper, DF}&\leq\frac{2}{|\mathcal V|}\int_{\mathbf v\in\mathcal V}\sigma_x^2-\bC_{x(\bv)\bar\by}(\bC_{\bar\by\bar\by}+\bC_{\tilde\by\tilde\by})^{-1}{\bC_{x(\bv)\bar\by}}^Td\bv\\
&=2\!\left[\sigma_x^2\!-\!\tr\!\left(\bTheta\bD_\balpha(\bC_{\bar\by\bar\by}\!+\!\bC_{\tilde\by\tilde\by})^{-1}\bD_\balpha\right)\right]\!\triangleq\bar\xi_{\rm approx, DF},\label{eq:xi_approx}
\end{align}
where
\begin{align}
\bTheta&\triangleq\frac{1}{|\calV|}\int_{\bv\in\calV}\bC_{x(\bv)\bm}^T\bC_{x(\bv)\bm}d\bv\\
&\approx\frac{1}{|\calV|}\int_{\bv\in\calV}\bC_{x(\bv)\bx}^T\bC_{x(\bv)\bx}d\bv=\bPhi,
\end{align}
and $\bPhi$ is defined in \eqref{eq:Phi_ij}. Moreover, using Lemma \ref{lemma:Cauchy_Schwarz} and the procedure similar to that of \eqref{eq:avMSEupper_AF_eigen}-\eqref{eq:avMSEupper2}, the average MSE can be further upper-bounded by
\begin{align}
\bar\xi_{\rm upper, DF}&\leq2\left[\sigma_x^2-\frac{\left(\tr\left(\bPhi\bD_\balpha\bD_\balpha\right)\right)^2}{\tr\left(\bPhi\bD_\balpha(\bC_{\bar\by\bar\by}+\bC_{\tilde\by\tilde\by})\bD_\balpha\right)}\right].
\end{align}
}

\section{Derivation of $\{\bC_{\tilde\by\tilde\by}\}_{i,j}$:}\label{APP:C_tilde_y}

Let us define $\tilde b_{i,k}\triangleq b_{i,k}-\hat b_{i,k}$ and $\Omega_i\triangleq\{o_i=1\}$ as the event that $o_i=1$ for $i=1,\ldots,M$. The diagonal element of $\bC_{\tilde\by\tilde\by}$ can be written as
\begin{align}
E[\tilde y_i^2]&=E\left[\left(\sum_{k=1}^{B_i}{(b_{i,k}- \hat b_{i,k})}2^{B_i-k}\Delta_{Q_i}\right)^2o_i^2\right]\\
&=2^{2B_i}\Delta_{Q_i}^2\left(\sum_{k=1}^{B_i}E\left[E\left[\tilde b_{i,k}^2o_i^2|o_i\right]\right]2^{-2k}+\sum_{k=1}^{B_i}\sum_{l\neq k}E\left[E\left[\tilde b_{i,k}\tilde b_{i,l}o_i^2|o_i\right]\right]2^{-k}2^{-l}\right)\\
&=2^{2B_i}\Delta_{Q_i}^2\left(\sum_{k=1}^{B_i}E\left[\tilde b_{i,k}^2|\Omega_i\right]\Pr(\Omega_i)2^{-2k}+\sum_{k=1}^{B_i}\sum_{l\neq k}E\left[\tilde b_{i,k}\tilde b_{i,l}|\Omega_i\right]\Pr(\Omega_i)2^{-k}2^{-l}\right).\label{eq.E_y_tilde_square}
\end{align}
By adopting BPSK modulation with transmit power $e_i/B_i$ per bit, the bit error probability of the sensor located in $\calV_i$ is
\begin{align}
q_i\!\triangleq\!\Pr(\hat b_{i,k}\!=\!1|b_{i,k}\!=\!0,\Omega_i)
\!=\!\Pr(\hat b_{i,k}\!=\!0|b_{i,k}\!=\!1,\Omega_i),
\end{align}
for $k=1,\ldots,B_i,~i=1,\ldots,M$. The expectation inside the first term of \eqref{eq.E_y_tilde_square} is
\begin{align*}
E\left[\tilde b_{i,k}^2\Big|\Omega_i\right]
&=1\cdot\Pr(b_{i,k}\neq\hat b_{i,k}|\Omega_i)+0\cdot\Pr(b_{i,k}=\hat b_{i,k}|\Omega_i)\\
\notag&=\Pr(\hat b_{i,k}=0|b_{i,k}=1,\Omega_i)\Pr(b_{i,k}=1|\Omega_i)+\Pr(\hat b_{i,k}=1|b_{i,k}=0,\Omega_i)\Pr(b_{i,k}=0|\Omega_i)\\
& =q_i.
\end{align*}
The expectation inside the second term of \eqref{eq.E_y_tilde_square} is
\begin{align*}
E\left[\tilde b_{i,k}\tilde b_{i,l}|\Omega_i\right]&=\!\Pr(\tilde b_{i,k}\!=\!1,\tilde b_{i,l}\!=\!1|\Omega_i)\!+\!\Pr(\tilde b_{i,k}\!=\!-1,\tilde b_{i,l}\!=\!-1|\Omega_i)\\
&~~~-\!\Pr(\tilde b_{i,k}\!=\!1,\tilde b_{i,l}\!=\!-1|\Omega_i)\!-\!\Pr(\tilde b_{i,k}\!=\!-1,\tilde b_{i,l}\!=\!1|\Omega_i)\\
\notag&=q_i^2[\Pr(b_{i,k}\!=\!1,b_{i,l}\!=\!1|\Omega_i)\!+\!\Pr(b_{i,k}\!=\!0,b_{i,l}\!=\!0|\Omega_i)\\
&~~~-\Pr(b_{i,k}\!=\!1,b_{i,l}\!=\!0|\Omega_i)\!-\!\Pr(b_{i,k}\!=\!0,b_{i,l}\!=\!1|\Omega_i)]\\
\notag&=q_i^2[1\!-\!2(\Pr(b_{i,k}\!=\!1,b_{i,l}\!=\!0|\Omega_i)\!+\!\Pr(b_{i,k}\!=\!0,b_{i,l}\!=\!1|\Omega_i))],
\end{align*}
where the last equality follows from properties of the natural binary code. By substituting the above into \eqref{eq.E_y_tilde_square}, we get
\begin{align}
E[\tilde y_i^2]&=2^{2B_i}\Delta_{Q_i}^2\left(\sum_{k=1}^{B_i}\alpha_iq_i2^{-2k}+q_i^2\eta_{i,j}\alpha_i\right)\\
&=2^{2B_i}\Delta_{Q_i}^2q_i\left(\frac{1-4^{-B_i}}{3}+q_i\eta_{i,j}\right)\alpha_i
\end{align}
The off-diagonal elements in $\bC_{\tilde\by\tilde\by}$ can be obtained similarly.

\section{Evaluation of $\pi_{\{i,k\},\{j,l\}}$:}\label{APP:pi}

To evaluate $\pi_{\{i,k\},\{j,l\}}$ in \eqref{eq:pi}, the joint probabilities $\Pr(b_{i,k}=1,b_{j,l}=0|\Omega_i,\Omega_j)$ and $\Pr(b_{i,k}=0,b_{j,l}=1|\Omega_i,\Omega_j)$ are required. To do this, let us define
\begin{align}
\calI_{i,k}^z\triangleq&\{\tilde x_i\in\mathbb R|b_{i,k}=z\}\\
\simeq&\bigcup_{\substack{b_{i,1},\ldots, b_{i,k-1}\\
\in\{0,1\}, b_{i,k}=z}}\left[\left(\sum_{t=1}^k b_{i,t}2^{B_i-t}-2^{B_i-1}\right)\Delta_{Q_i},\left(\sum_{t=1}^k b_{i,t}2^{B_i-t}\!+\!2^{B_i-k}-2^{B_i-1}\!\right)\!\!\Delta_{Q_i}\!\right]
\end{align}
as the set of measurement values $\tilde x_i$ at sensor $i$ that yields $b_{i,k}=z$, where $z\in\{0,1\}$. The approximation follows the fact that $\Pr(|\tilde x_i|>W)\approx0$. Notice that $\calI_{i,k}^z$ can be written as the union of $2^{k-1}$ disjoint intervals. Therefore, for $i=j$ and $k<l$, we have
\begin{align}
\Pr(b_{i,k}=1,b_{j,l}=0|\Omega_i,\Omega_j)
&=\Pr(\tilde x_i\in\calI_{i,k}^1\cap\calI_{i,l}^0|\Omega_i)\\
&=\sum_{\substack{b_{i,1},\ldots, b_{i,k-1},\\
b_{i,k+1},\ldots, b_{i,l-1}\in\{0,1\}\\
b_{i,k}=z_1, b_{i,l}=z_2}}\left[Q\left(\frac{\left(\sum_{\substack{t=1}}^{l} b_{i,t}2^{B_i-t}-2^{B_i-1}\right)\Delta_{Q_i}}{\sqrt{\sigma_x^2+\sigma_n^2}}\right)\right.\notag\\
&~~~~\left.-Q\left(\frac{\left(\sum_{t=1}^{l} b_{i,t}2^{B_i-t}+ 2^{B_i-l}-2^{B_i-1}\right)\Delta_{Q_i}}{\sqrt{\sigma_x^2+\sigma_n^2}}\right)\right].
\end{align}
The result is similar for $k>l$. Moreover, since $\tilde x_i$ and $\tilde x_j$, for $i\neq j$, are jointly Gaussian random variables with mean $0$, variance $\sigma_{\tilde x}^2= \sigma_x^2+\sigma_n^2$, and correlation coefficient $\tilde \rho(\bv_i, \bv_j)=\frac{\sigma_x^2}{\sigma_x^2+\sigma_n^2}\rho(\bv_i,\bv_j)$, we have, for $i\neq j$,
\begin{align}
\notag&\Pr(b_{i,k}=1,b_{j,l}=0|\Omega_i,\Omega_j)\\
&=\Pr(\tilde x_i\in\calI_{i,k}^1,\tilde x_j\in\calI_{j,l}^0)\\
&=\int_{\tilde x_j\in\calI_{j,l}^0}\int_{\tilde x_i\in\calI_{i,k}^1}\frac{e^{-\frac{\tilde x_i^2-2\tilde\rho(\bv_i,\bv_j)\tilde x_i\tilde x_j+\tilde x_j^2}{2\sigma_{\tilde x}^2(1-\tilde\rho^2(\bv_i,\bv_j))}}}{2\pi\sigma_{\tilde x}^2\sqrt{1-\tilde\rho^2(\bv_i,\bv_j)}}d\tilde x_id\tilde x_j\\
&\mathop{=}^{(a)}\int_{\tilde x_j\in\calI_{j,l}^0}\frac{1}{\sqrt{2\pi\sigma_{\tilde x}^2}}\left[\int_{x_i'\in\calI^{1'}_{i,k}}\frac{e^{-\frac{x_i'^2}{2}}}{\sqrt{2\pi}}dx_i'\right]e^{-\frac{\tilde x_j^2}{2\sigma_{\tilde x}^2}}d\tilde x_j,\\
\notag&\mathop{=}^{(b)}\!\!\!\sum_{\substack{b_{j,1},\ldots, b_{j,l-1}\\
\in\{0,1\}, b_{j,l}=0}}\!\!\!\int_{\!(\sum_{t=1}^{l}\! b_{j,t}2^{B_j-t}-2^{B_j-1})\Delta_{Q_j}}^{\left(\sum_{t=1}^{l}\! b_{j,t}2^{B_j\!-\!t}\!+2^{B_j\!-l}\!-2^{B_j\!-1}\!\right)\Delta_{Q_i}}\!\!\!\!\!\!\!\!\!\!\sum_{\substack{b_{i,1},\ldots, b_{i,k-1}\\
\in\{0,1\}, b_{i,k}=1}}\!\!\!\!\frac{e^{-\frac{\tilde x_j^2}{2\sigma_{\tilde x}^2}}}{\sqrt{2\pi\sigma_{\tilde x}^2}}\\
\notag&\quad\left[Q\!\left(\frac{\left(\sum_{t=1}^{k} b_{i,t}2^{B_i-t}-2^{B_i-1}\right)\Delta_{Q_i}-\tilde\rho(\bv_i,\bv_j)\tilde x_j}{\sqrt{\sigma_{\tilde x}^2\left(1-\tilde\rho^2(\bv_i,\bv_j)\right)}}\!\right)-\right.\\
&\quad\left.Q\!\left(\!\frac{\left(\!\sum_{t=1}^{k} b_{i,t}2^{B_i-t}\!+\!2^{B_i-k}\!-2^{B_i-1}\!\!\right)\!\Delta_{Q_i}\!\!-\!\tilde\rho(\bv_i,\bv_j)\tilde x_j}{\sqrt{\sigma_{\tilde x}^2\left(1\!-\!\tilde\rho^2(\bv_i,\bv_j)\right)}}\!\right)\!\right]\!\!d\tilde x_j
\end{align}
where (a) follows from the change of variable
\begin{align*}
x_i'=\frac{\tilde x_i\!-\!\tilde\rho(\bv_i,\bv_j)\tilde x_j}{\sqrt{\sigma_{\tilde x}^2(1\!-\!\tilde\rho^2(\bv_i,\bv_j))}}
\end{align*}
with $\calI^{1'}_{i,k}\triangleq\big\{\frac{\tilde x_i-\tilde\rho(\bv_i,\bv_j)\tilde x_j}{\sqrt{\sigma_{\tilde x}^2(1-\tilde\rho^2(\bv_i,\bv_j))}}\in\mathbb R\big|b_{i,k}=1\big\}$ and (b) follows from the definition of $\calI_{j,l}^0$ and $\calI^{1'}_{i,k}$.
The probability $\Pr(b_{i,k}=0,b_{j,l}=1|\Omega_i,\Omega_j)$ can be evaluated similarly.

\end{document}